\documentclass[12pt,table]{article}
\usepackage[utf8]{inputenc}
\usepackage[english]{babel}
\usepackage{amsfonts,amssymb,amsmath,amsthm}
\usepackage{dsfont}
\usepackage{bm}
\usepackage{longtable}
\usepackage{authblk}
\usepackage{multicol}
\usepackage{multirow}
\usepackage{eurosym}
\usepackage{natbib}
\usepackage{eqnarray}
\usepackage{makecell}
\usepackage{appendix}
\usepackage{xcolor}
\usepackage{adjustbox}
\usepackage[colorlinks = true,
            linkcolor = blue,
            urlcolor  = blue,
            citecolor = blue,
            anchorcolor = blue]{hyperref}
\usepackage{subcaption}
\usepackage{tabu}
\usepackage{soul}
\usepackage[paper=a4paper,margin=1.2in]{geometry}
\usepackage{setspace}
\newcommand{\E}{\mathbb{E}}
\newcommand{\Real}{\mathbb{R}}

\newcommand\widebar[1]{\mathop{\overline{#1}}}

\makeatletter
\renewcommand{\maketitle}{\bgroup\setlength{\parindent}{0pt}
\begin{flushleft}
  {\LARGE\textbf{\@title}} \vspace{0.3cm} \newline
  \@author  \vspace{0.9cm} \hrule \vspace{0.6cm}
\end{flushleft}\egroup
} 
\makeatother

\renewenvironment{abstract}
 {\small
  \begin{center}
  \bfseries \abstractname\vspace{-.5em}\vspace{0pt}
  \end{center}
  \list{}{%
    \setlength{\leftmargin}{0cm}
    \setlength{\rightmargin}{\leftmargin}%
  }%
  \item\relax}
 {\endlist}
 
\title{Outlier detection in network revenue management}

\author[a]{\small Nicola RENNIE}
\author[b,*]{\small Catherine CLEOPHAS}
\author[c]{\small Adam M. SYKULSKI}
\author[d]{\small Florian DOST}

\affil[a]{\footnotesize STOR-i Centre for Doctoral Training, Lancaster University, LA1 4YW, UK. (\href{mailto:n.rennie@lancaster.ac.uk}{n.rennie@lancaster.ac.uk})}
\affil[b]{\footnotesize Institute of Business, Christian-Albrechts-University Kiel, Kiel, Germany. (\href{mailto:cleophas@bwl.uni-kiel.de}{cleophas@bwl.uni-kiel.de})}
\affil[c]{\footnotesize Dept. of Mathematics, Imperial College London, SW7 2AZ, UK. (\href{mailto:adam.sykulski@imperial.ac.uk}{adam.sykulski@imperial.ac.uk})}
\affil[d]{\footnotesize Institute of Business and Economics, Brandenburg University of Technology, 03046 Cottbus, Germany. (\href{mailto:florian.dost@b-tu.de}{florian.dost@b-tu.de})}
\affil[*]{Corresponding Author}

\date{\vspace{-8ex}}

\begin{document}

\maketitle

\begin{abstract}
This paper presents an automated approach for providing ranked lists of outliers in observed demand to support analysts in network revenue management. Such network revenue management, e.g. for railway itineraries, needs accurate demand forecasts.  However, demand outliers across or in parts of a network complicate accurate demand forecasting, and the network structure makes such demand outliers hard to detect.

We propose a two-step approach combining clustering with functional outlier detection to identify outlying demand from network bookings observed on the leg level. The first step clusters legs to appropriately partition and pool booking patterns. The second step identifies outliers within each cluster and uses a novel aggregation method across legs to create a ranked alert list of affected instances. Our method outperforms analyses that consider leg data without regard for network implications and offers a computationally efficient alternative to storing and analysing all data on the itinerary level, especially in highly-connected networks where most customers book multi-leg products. A simulation study demonstrates the robustness of the approach and quantifies the potential revenue benefits from adjusting demand forecasts for offer optimisation. Finally, we illustrate the applicability based on empirical data obtained from Deutsche Bahn.

\end{abstract}


\section{Introduction and State of the Art} 
\label{sec:introduction}

Revenue management (RM) is a challenging task for network service providers. The concept entails controlling the set of product offers over a fixed sales horizon such that, given the predicted demand for the offers, the expected revenue from selling a limited capacity is maximal. To thus maximise revenue, the firm has to forecast the expected demand for all products that require capacity on mutual resources. Examples include transport itineraries that cross several network legs, and hospitality offers that combine room availability for multiple nights. 

Several existing contributions, e.g. \citet{Weatherford2002} and \citet{Rennie2021}, demonstrate the negative effects of inaccurate demand forecasts on revenue performance but neglect network effects. Motivated by this, we propose a new approach to detect outliers in network bookings, thereby supporting forecast corrections for improved network revenue management.

\subsection{Terminology}
For simplicity, we employ a transport-based terminology throughout this paper: a \emph{leg} describes a direct non-stop connection between two stations in a network, and an \emph{itinerary} is any combination of legs that can be jointly booked as one product. A \emph{departure} describes a journey along a connected series of legs that leaves the origin station at a unique time and date. 

We denote the accumulation of \emph{bookings} across the sales horizon as a \emph{booking pattern}. We define an \emph{outlier} as a booking pattern resulting from short-term systematic demand changes for one or several related network itineraries. These outliers occur when demand deviates from the baseline due to unforeseen events. For example, demand increases for a specific destination affect the entire itinerary. In consequence, deviations in the booking patterns are observable on the legs arriving at that destination and in the feeder legs. 

Capacity-based revenue management differentiates offers through  \emph{fare classes}. Fare classes describe combinations of fares and tariffs at which the firm offers a product. Customers booking a ticket for a specific departure may choose from several offered fare classes. For instance, the cheapest offer could be fare class `M', costing 20 Euros and entailing a no-refund tariff. 

\subsection{Existing Work}

RM is a well-studied problem for many different products and services \citep{Talluri2004}. Still, only recently the specific issues around network services and demand forecasting have come into focus. E.g. \citet{Klein2020} review how single-leg practices to RM generalise to the network setting. \citet{Weatherford2016} surveys RM forecasting methods and focuses on airline itinerary-level forecasting. 

So far, few authors have examined demand outliers in RM data. For historical hotel booking data, \citet{Weatherford2003} discuss a simple method of removing observations that are more than $\pm 3\sigma$ away from the mean. \citet{Rennie2021} apply functional analysis to detect outliers on individual legs. Neither, however, consider outliers affecting multiple legs of a network. In \citet{azadeh2013railway}, the authors identify outliers in network railway bookings via a simple rule to remove them before forecasting future demand. For a slightly different perspective, \citet{kumar2020evaluating} analyse transit demand for outliers to detect special events. Notably, existing research on outlier detection frequently focuses on  binary outlier detection without regard for quantifying how critical an outlier is. 
 
Practical network RM relies on manual forecast adjustments \citep{quante2009revenue, schutze2020revenue}. Previous research has shown that the resulting judgemental forecasts can be biased and even superfluous \citep{lawrence2006, DeBaets2020}. \citet{Perera2019} note that forecasting support tools can improve user judgement by reducing complexity for the analyst. Analysts' time is limited, so they cannot investigate every departure flagged as an outlier. For example, Deutsche Bahn experts estimate that they can reasonably adjust less than 1\% of forecasts. Therefore, ranking outliers by criticality is crucial.

Beyond RM, \citet{barrow2018impact} propose a functional approach for outlier detection in call arrival forecasting without regard for network effects. General outlier detection in networks often focuses on identifying outlying parts of the network. \citet{Fawzy2013} use this approach in wireless sensor networks to find faulty nodes. \citet{Ranshous2015} consider the extension to identify outlying nodes when the network changes over time. Most research on dynamic networks concentrates on analysing a single time series connected to each node rather than a set of time series, as required when booking patterns are reported for multiple departures. \citet{Hyndman2016} note that the problem of identifying unusual time series within a collection is not as extensively studied as other outlier detection problems. In this paper, we benchmark the approach suggested by \citet{Hyndman2016}, which employs principal component analysis (PCA), against our newly proposed approach.

\subsection{Contribution}

We shall study booking patterns that result when customers book not just a single resource (leg) but network products that require multiple resources (itineraries). Such booking patterns may be reported on the leg or the itinerary level -- in this paper we assume that they are reported per leg and departure. This applies in the case of Deutsche Bahn, which serves as a motivation and empirical demonstration for the work presented here.

Network effects challenge outlier detection in two ways: On the one hand, demand outliers on the itinerary level affect bookings on all legs included in the itinerary. On the other hand, such outliers may not be recognisable when only considering leg bookings independently, given the noise from other itineraries overlapping those legs. As a result, directly extracting outliers from booking data collected in an entire, realistically sized network is likely an intractable problem. To circumvent this problem, in this paper, we aggregate and analyse booking patterns from \emph{legs} instead of \emph{itineraries}, as this allows for computationally and statistically tractable network-wide outlier detection. In Section~\ref{sec:conclusion}, we further discuss the choice of leg-level vs itinerary-level-based analysis and point out how our procedures could be adapted to itinerary-level outlier detection.

Our network outlier detection procedure: (i) clusters legs with similar booking patterns and (ii) detects joint outliers within each cluster to compile ranked alert lists of outlying departures and affected legs. Our methodology significantly improves outlier detection performance in a network setting versus alternative methods.

In more detail, our proposed approach first clusters legs by measuring the similarity of booking patterns via functional dynamical correlation \citep{Dubin2005}. We suggest this measure for its freedom from restrictive assumptions. As the proposed approach is modular, other correlation measures could be used for the same end. In the second step, the proposed approach detects outliers from booking patterns within each cluster by combining the functional data analysis methods of \citet{Febrero2008, Hubert2012, Rennie2021} with a novel within-cluster aggregation, which generates a ranked alert list of outliers using extreme value theory. This alert list can help analysts to identify the need for further analysis and adjustments. We consider an outlier as more critical if it indicates a larger demand shift and if it is identified across multiple legs. Factors such as the average fare on legs where outliers are detected, or the revenue at risk from faulty forecasts could also be incorporated into the definition of an outlier's criticality.

Finally, analysts have several choices when tasked with forecast adjustment for network services. The best choice is not obvious, and we further quantify the impact of different potential adjustments on revenue in a simulation study, following concepts outlined in \citet{kimms2007simulation}.

In summary, this paper contributes (i) a method for identifying network legs that will benefit from joint outlier detection and (ii) a method to aggregate outlier detection across any number of legs to create a ranked alert list. To thoroughly evaluate the proposed approach, we offer (iii) wide-ranging simulation studies to benchmark the method's outlier detection performance against and to quantify the potential revenue improvements from forecast adjustments and (iv) a demonstration of applicability on empirical railway booking data from Deutsche Bahn.


\section{Method} 
\label{sec:method} 

Several network products may rely on common resources when demand concerns multiple legs at once -- in the transport example, even passengers that booked different itineraries often have to traverse the same legs. Therefore, specific legs share common outliers, as, for example, a sudden increase in demand from passengers travelling from one end of the network to attend an event at the other end would increase demand for each of the in-between legs. Neither considering each leg independently, nor jointly considering the whole network, will create the best results when the network spans multiple regions that differ strongly in demand -- see Section \ref{sec:outlier_results} and Appendix \ref{app:subset_itin_outliers}. This raises the question of which legs to consider jointly for outlier detection. 

To find an answer, in Section \ref{sec:clustering}, we adapt a method by \cite{Zahn1971} to \emph{cluster} legs such that (i) legs in the same cluster share demand and can be considered jointly for outlier detection, and (ii) legs in different clusters experience distinct demand and should be considered separately. Subsequently, in Section \ref{sec:network_outliers}, we suggest a method for analysing bookings within one such cluster. Based on this, we propose a method to rank departures by the severity of identified outliers.

\subsection{Clustering legs using correlation-based minimum spanning trees} \label{sec:clustering}

\begin{figure}[ht]
    \centering
    \includegraphics[width=\textwidth]{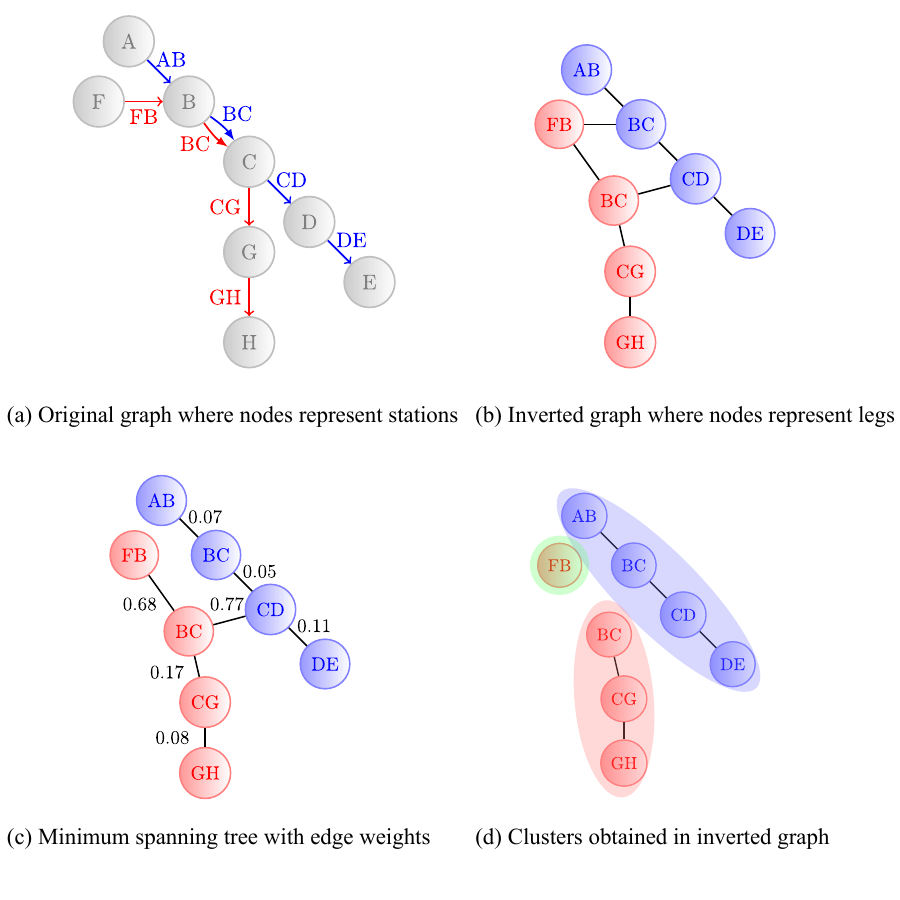}
    \caption{Correlation-based minimum spanning tree clustering}
    \label{fig:network_method}
\end{figure}

To cluster legs based on correlations in observed bookings, we first consider the network as a graph where nodes represent the stations and edges represent the legs of a journey. Figure \ref{fig:network_method}a illustrates this on a simple network. To illustrate the concept, we rely on an example from the transport domain: In this example, two train lines (red and blue) intersect at two stations (B and C). The red train arrives at stations B and C before the blue train, which creates two possible transfer connections for passengers: (i) switch from red to blue at B, (ii) switch from red to blue at C. Transfers from the blue to red train are not feasible.

Standard graph clustering algorithms, as exemplified in \citet{Schaeffer2007}, seek to cluster the nodes of the graph. In contrast, we wish to cluster similar edges, which correspond to legs in the railway example (Figure \ref{fig:network_method}a). Hence, we invert the graph to make existing clustering algorithms applicable. In this inversion (Figure \ref{fig:network_method}b), the directed edges become nodes, e.g. the edge from A to B becomes node AB. The inverted graph features an undirected edge between two nodes when:
\begin{itemize}
    \item both legs are in the same train line and share a common station, e.g. legs CD and DE are connected through station D, or
    \item the legs are in different train lines but share a common transfer station where a connection is possible, e.g. leg FB (red line) and BC (blue line) are connected through station B. However, AB (blue line) and BC (red line) would not be connected by an edge as no connection can be made between them (as we have assumed the red train arrives at B and C before the blue train).
\end{itemize}
In theory, this transformation could also create edges between legs that share a common entry or exit node, e.g. FB (red line) and AB (blue line), or CG (red line) and CD (blue line). Given that such pairs of legs would never occur in the same itinerary, we would not expect demand outliers to affect both legs. Therefore, inserting an edge between them, potentially allowing them to be in the same cluster, is counter-intuitive. In addition, exploratory analyses of the empirical data found that correlations between these types of legs were very low across the entire network section.

The algorithm aims to assign those legs that experience \textit{similar bookings} to the same cluster and those that experience \textit{dissimilar bookings} to distinct clusters. A corresponding metric only needs to consider the similarity between adjacent legs that share a connecting station since edges do not otherwise exist in the inverted graph. We propose to quantify this similarity via the correlation between booking patterns. 

\emph{To calculate correlations between booking patterns,} we compute the functional dynamical correlation \citep{Dubin2005}. Functional dynamical correlation is based on calculating scalar products between pairs of smoothed booking patterns; the appendix provides further details. We use the average of these paired correlations over time as the similarity measure between two legs. Unlike more common statistical correlation measures, such as Pearson correlation, functional dynamical correlation does not assume a specific type of relationship between variables (e.g. linearity). It also accounts for the time dependency between observations within the booking horizon when the intervals between observations vary. For example, in the empirical RM data analysed in Section \ref{sec:empirical}, the time between observations decreases as the departure date approaches. Further, alternative measures for calculating correlations from functional data (such as functional canonical correlation) often make restrictive assumptions, which real data does not fulfil \citep{He2003}.

We benchmark the clustering algorithm under alternative correlation measures in Appendix \ref{app:clust_benchmark}.

\emph{To represent the relationship between legs in the network,} i.e. the nodes in the inverted graph, we attach weights to the edges in the inverted graph. These weights are interpreted as distances: a higher edge weight indicates that the connected nodes are more dissimilar. Therefore, an applicable weight function should be non-negative. Further, the weight function needs to ensure that any negatively correlated legs are marked as more dissimilar. Even though a negative correlation may imply that outlier demand jointly affects both legs, we expect it to affect negatively correlated legs differently. Therefore, these require different adjustments from an analyst and should be in different clusters. To satisfy these requirements, we define the edge weights as:
\begin{equation}
    w_{(ij, jk)} = 1 - \rho(ij, jk),
\end{equation}
where $\rho(ij, jk)$ is the correlation between bookings on legs $ij$ and $jk$. Though the use of functional dynamical correlation as a measure of similarity between time series is not new, its application as an edge weight in a network setting, to our knowledge, is novel.

\emph{To allow for irregular cluster shapes,} we recommend a minimum spanning tree (MST) algorithm   \citep{Prim1957}. For example, in Figure \ref{fig:network_method}b, a cluster may include AB and DE because they are in the same line, rather than clustering AB and FB. Minimum spanning tree approaches work well for clusters with irregular boundaries \citep{Zahn1971}. Alternative clustering approaches (such as $k$-means) often assume a specific shape of clusters (spherical, for $k$-means). MST-based clustering approaches also do not assume that clusters are of similar sizes \citep{Peter2010}. This makes them particularly suitable for transportation networks constructed as a series of interlocking lines, where the points of intersection are often not equally spaced. For example, MST-based approaches have previously been used in optimising layouts of railway networks \citep{Liang2020}. 

A \textit{spanning tree} of a graph is a subgraph that includes all vertices in the original graph and a minimum number of edges, such that the spanning tree is connected. Then, the MST is the spanning tree with the minimum summed edge weights -- see Figure \ref{fig:network_method}c. Since the inverted graph is weighted, we use Prim's algorithm \citep{Prim1957} to calculate the MST -- Appendix \ref{app:prims} provides a detailed introduction. Any one-to-one transformation of the weight function, $w_{(ij, jk)}$ will produce an identical minimum spanning tree.

There are two approaches to obtaining clusters from an MST: (i) pre-defining the number of clusters as $k$ and removing the $k-1$ edges with the highest weight; or (ii) setting a threshold for the edge weights and removing all edges with weights above some threshold, creating an emergent number of clusters. Here, we implement the threshold-based approach, ensuring that each cluster has the same minimum level of correlation. In contrast, setting the number of clusters in advance could result in very heterogeneous levels of correlation across clusters. Further, setting $k$ too low may result in legs with dissimilar features being grouped together. We apply a threshold correlation of 0.5 -- the level at which legs are more correlated than they are not. This corresponds to a transformed edge weight of 0.5. In the example given in Figure \ref{fig:network_method}c, this means removing all legs with a weight above 0.5, resulting in the three clusters shown in Figure \ref{fig:network_method}d. The choice of this clustering threshold will impact the number of alert lists produced. Therefore, we recommend considering factors such as staffing resources and any current (informal) network clustering when choosing this threshold. 

\emph{While the outlier detection procedure described next applies to individual clusters, it does not require a particular clustering approach.} Hence, other implementations may employ alternative approaches, as reviewed in \citet{Schaeffer2007}. In particular, depending on the business context of the network service, alternative clustering algorithms may be more appropriate. The network topology should drive the choice of which clustering algorithm is most appropriate: That topology may differ, e.g., when considering airlines versus bike rentals versus railways, as discussed in \citet{Rennie2022}. The choice of an MST-based approach, which often returns linear clusters, is appropriate for the railway application motivating the paper at hand, given the linear nature of the underlying network structure. We further evaluate the performance of MST clustering in this application in Appendix \ref{app:clust_benchmark}.

Furthermore, edge-based clustering could replace the graph inversion and node-based clustering presented here. However, literature on edge-based clustering is far more limited, and such approaches tend to improve the visualisation of networks with a very high number of edges by reducing the number of edge crossings rather than grouping together the most similar edges \citep{Qu2007}. In contrast, inversion and node-based clustering aim to group network legs that exhibit the highest degree of similarity. However, alongside these advantages, there may be some drawbacks. The node-based approach requires deciding on criteria to select edges to include in the inverted graph.

\subsection{Detecting outliers in clusters of legs} \label{sec:network_outliers}
Given established clusters, we propose identifying demand outliers within each cluster and quantifying their severity to provide a {\em ranked alert list} of departures. The previously described clustering allows for processing the outlier detection in parallel for separate clusters, enabling efficient computing.

To identify which departures to include in the alert list, we consider the {\em functional  depth} of the booking patterns, as in \citet{Rennie2021}. This step could also rely on other measures of exceedance, including univariate ``threshold" approaches, which look at aggregated bookings and ignore the distribution of bookings over time. We propose to rely on functional depth, as previous work has found this to be the most effective as an outlier detection mechanism \citep{Rennie2021}.

To compute the functional depth, consider $N$ departures observed over $L$ legs. Let \(\bm{y}_{nl} =\left(y_{nl}(t_1), \hdots, y_{nl}(t_{T}) \right)\) be the booking pattern for the $n^{th}$ departure on leg $l$, observed over $T$ booking intervals $t_1,\ldots,t_T$. Let $\mathcal{Y}_l$ be the set of $N$ booking patterns for leg $l$. For each leg and departure, calculate the functional depth  ($d_{nl}$) given the related booking patterns following the approach given in \citet{Hubert2012} and detailed in Appendix \ref{app:func_depth}. The functional depths take on positive values, with smaller values of the depths relating to more outlying booking patterns.

For each leg $l$, we calculate a threshold for the functional depth using the approach of \citet{Febrero2008}. This method (i) resamples the booking patterns with probability proportional to their functional depths (such that any outlying patterns are less likely to be resampled), (ii) smooths the resampled patterns, and (iii) sets the threshold \(C_l\) as the median of the \(1^{st}\) percentiles of the functional depths of the resampled patterns. Here, we use the \(1^{st}\) percentile of the depths as the default threshold, as this has been found to work well in practice \citep{Febrero2008, Rennie2021}. Booking patterns with a functional depth below the threshold $C_l$ are classed as outliers. We explore alternative threshold choices in Appendix \ref{app:threshold_results}. 

To create ranked alert lists, we first define $z_{nl}$ to be the normalised difference between the functional depth and the threshold: 
\begin{equation} \label{eqn:z_nl}
    z_{nl} = \frac{C_l - d_{nl}}{C_l}.
\end{equation}
This transforms the depth measure $d_{nl}$ into a measure of \textit{threshold exceedance}. Values of $z_{nl}$ greater than zero relate to booking patterns classified as outliers. Normalising by the threshold, $C_l$, ensures the values of $z_{nl}$ are comparable between different legs. 

\begin{figure}[!ht]
	\centering
		\includegraphics[width=\textwidth]{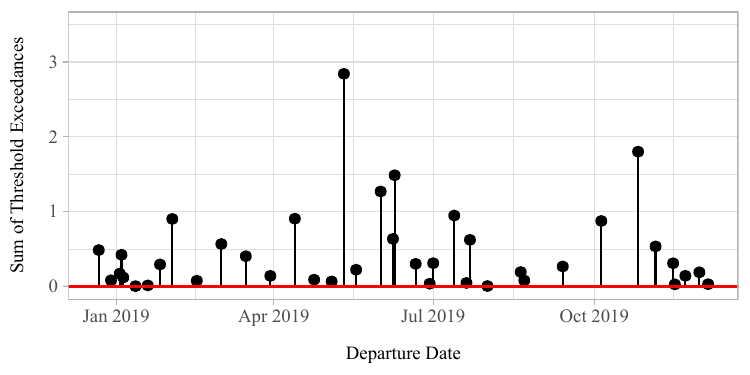} 
	\caption{$z_{n}$ as defined in equation~\eqref{eq:z_n} for a four leg section of the Deutsche Bahn network} 
	\label{fig:zn}
\end{figure}

Next, we define the sums of threshold exceedances across legs:
    \begin{equation} \label{eq:z_n}
        z_n = \sum_{l=1}^{L} z_{nl} \mathds{1}_{\{z_{nl} > 0\}}.
    \end{equation}
We sum only those values of $z_{nl}$ that are greater than zero to avoid outliers being masked when they occur only in a subset of legs. This sum implicitly accounts for both the size of an outlier -- larger outliers further exceeding the threshold, resulting in larger values of $z_{nl}$ -- and for the number of legs where a departure is classified as an outlier (by summing a larger number of non-zero values). To provide an example, Figure \ref{fig:zn} shows those values of $z_n$ that exceed zero for a four-leg section of the Deutsche Bahn network as discussed further in Section \ref{sec:db_results}. These values of $z_n$ correspond to departures where the booking pattern for {\em at least} one leg is identified as an outlier. In contrast, all other departures have no detected outliers in any leg such that $z_n=0$.

To create a ranked list of outlier departures, i.e. those with a non-zero-sum of threshold exceedances, we assign a severity $\theta_n$. A higher value of $\theta_n$ indicates the departure is more likely to be affected by extreme outlier demand and hence should be targeted first by RM analysts. 
 
To model threshold exceedances, we turn to extreme value theory (EVT) -- a branch of statistics that deals with modelling rare events occurring in the tails of a distribution. Given that outliers are unusual events, which occur in the tails of distributions, EVT is a clear direction to turn to for modelling outliers -- see \citet{Hyndman2019}. There are two common approaches to EVT: (i) block maxima, which examines the maximum value in evenly-spaced blocks of time, e.g. annual maxima, and (ii) peaks over the threshold, which examines all observations that exceed some threshold \citep{Leadbetter1991}. The generalised Pareto distribution (GPD) is commonly used to model the tails of distributions in the peaks over threshold approach \citep{Pickands1975}. Motivated thus, we fit a generalised Pareto distribution (GPD) to the sum of threshold exceedances given in equation~\eqref{eq:z_n}. The GPD has three parameters with probability density function:
\begin{equation}
    f(x\vert\mu, \sigma, \xi) = \frac{1}{\sigma} \left(1 + \frac{\xi(x-\mu)} {\sigma}^{\left(-\frac{1}{\xi} - 1\right)} \right),
\end{equation}
for
\begin{equation}
    x \in  \begin{cases}
               [\mu, \infty) & \xi \geq 0 \\
               [\mu, \mu - \frac{\sigma}{\xi}]  & \xi < 0.
            \end{cases}
\end{equation}
Here, \(\mu\) specifies the location, \(\sigma\) the scale, and \(\xi\) the shape of the distribution. We fit the parameters using maximum likelihood estimation \citep{Grimshaw1993} via the R package \texttt{POT} \citep{pot}. A kernel density estimate of the empirical distribution of \(z_n > 0\) from Figure \ref{fig:zn} is shown in Figure \ref{fig:zn_dist}a. The resulting fitted GPD is shown in Figure \ref{fig:zn_dist}b. As the further analysis in Appendix \ref{app:pp_plots} shows, the GPD fit appears reasonable compared to the empirical distribution. 

\begin{figure}[!ht]
    \centering
    \includegraphics[width=\textwidth]{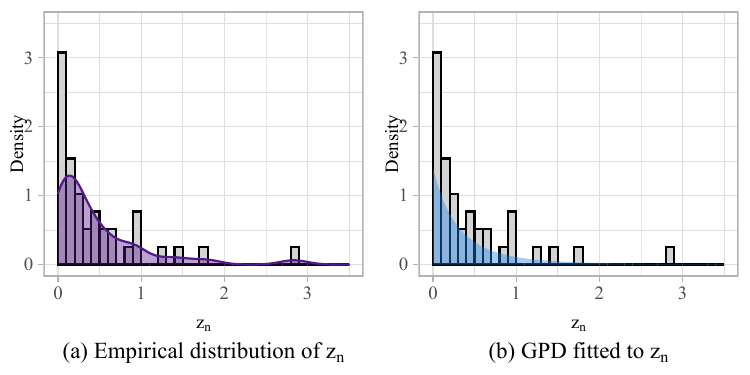}  
    \caption{Distribution of $z_n$ values from Figure \ref{fig:zn}}
    \label{fig:zn_dist}
\end{figure}

Two common issues arise in fitting GPDs: (i) the choice of threshold and (ii) the independence of the data points. When the threshold is too low, the assumption of a GPD no longer holds; when it is too high, there are too few data points to fit. We select a threshold of 0, i.e. we fit the GPD to values of \(z_n > 0\). Rather than change the threshold at the GPD level, we control the number of observations the GPD is fitted to by varying the percentile used for the individual leg thresholds, $C_l$. We choose $C_l$ as suggested by \citet{Febrero2008} and find that this choice works well and provides sufficient outlying points to fit a GPD in both simulated and empirical data. 

To account for the second issue, applications of extreme value theory frequently first \textit{decluster} the peaks over the threshold to ensure independence between observations \citep{Fawcett2007}. To that end, the analysis may only consider the maximum of two peaks within a small time window. For transport departures, it is theoretically possible that observed outliers may be dependent; e.g. increased demand caused by Easter affects not only Easter Sunday but also the surrounding days. However, similar outliers may also result from independent events. As we aim to identify outlying departures rather than the underlying events, this argument causes us not to decluster here. 

We define $\theta_n$ as the non-exceedance probability given by the CDF of the GPD: 
\begin{equation} \label{eqn:theta}
    \theta_n = F_{(\mu, \sigma, \xi)}(z_n) = \begin{cases}
               1 - \left(1 + \frac{\xi(z_n - \mu)}{\sigma} \right)^{-\frac{1}{\xi}} & \xi \neq 0 \\
               1 - \exp\left(-\frac{(z_n - \mu)}{\sigma}\right) & \xi = 0
            \end{cases}
\end{equation}
Formally, $\theta_n$ is the probability that, given an outlier occurs, the sum of threshold exceedances is at least as large at $z_n$. Thus, it is \textit{not} the probability that a departure is an outlier. However, we use this non-exceedance probability as a measure of outlier \textit{severity} on a scale of 0 to 1. 

Departures with functional depths that do not fall below the threshold on any legs carry a severity of zero, i.e. they are classified as regular departures. It is conceivable to estimate the uncertainty of $\theta_n$ \citep{Smith1985} to determine further levels of criticality, e.g., if there are several departures with the same outlier severity, the one with the smallest uncertainty would be ranked first. However, given the continuous nature of the data, it is unlikely that multiple departures carry an identical severity. Hence, we leave uncertainty estimation to future research.

From the severity defined in equation~\eqref{eqn:theta}, we construct a \emph{ranked alert list} containing all departures with a non-zero outlier severity. Although functional depth could be directly used to construct the ranked alert list, computing the severity provides a measure of the difference between ranks and is more easily interpreted by analysts. The top 8 ranked outliers relating to Figure \ref{fig:zn}, are shown in Table \ref{tab:alert_list}.
\begin{table}[!ht]
    \centering
    \begin{tabular}{cccc}
    \hline \hline
        \textbf{Ranking} & \textbf{Departure} & \textbf{Severity} & \textbf{Legs with $z_{nl} > 0$} \\ \hline
        1                & 11/05/2019                             & 0.985                & AB, BC, CD, DE            \\
        2                & 26/10/2019                             & 0.960                & AB, BC, CD, DE            \\
        3                & 09/06/2019                             & 0.942                & AB, BC, CD, DE            \\
        4                & 01/06/2019                             & 0.922                & AB, BC, CD, DE            \\
        5                & 13/07/2019                             & 0.874                & AB, BC, CD, DE            \\
        6                & 13/04/2019                             & 0.865                & CD, DE            \\
        7                & 02/02/2019                             & 0.864                & CD, DE            \\
        8                & 05/10/2019                             & 0.857                & AB, BC, CD, DE            \\
        $\vdots$         & \multicolumn{1}{c}{$\vdots$}           & $\vdots$             & $\vdots$                  \\
        \hline \hline
    \end{tabular}
    \caption{Ranked alert list for cluster $= \{AB, BC, CD, DE\}$}
    \label{tab:alert_list}
\end{table}

In practice, RM analysts' time and resources allow them to examine and adjust controls or forecasts only for a limited number of suspicious booking patterns. Those departures that (i) exceed the functional depth threshold in only one leg or (ii) exceed the threshold only to a small degree have lower but strictly non-zero severity. These outliers are most likely false positives and potentially waste analysts' time. Hence, we suggest limiting the length of the list in practice. 

To limit the length of the alert list, we might (i) only include departures if their severity is above some threshold or (ii) set a maximum length. Since we wish to control the number of alerts an analyst will receive, we analyse outlier detection performance as dependent on the maximum length of the alert list. Recall that we classify departures as outliers if and only if their outlier severity exceeds zero. Therefore, if the required length of the alert list exceeds the number of identified outliers, we do not include further departures. Appendix \ref{app:outliers_perc} features further results on the outlier detection performance when varying the outlier severity threshold. 


\section{Outlier Detection Performance}
\label{sec:simulation} 

We first implement a simulation study to evaluate the outlier detection performance given known outliers. By varying the demand for itineraries in one cluster, we create outliers that are observable on both the leg and network levels.

\begin{figure}[!ht]
    \centering
    \includegraphics[width=0.7\textwidth]{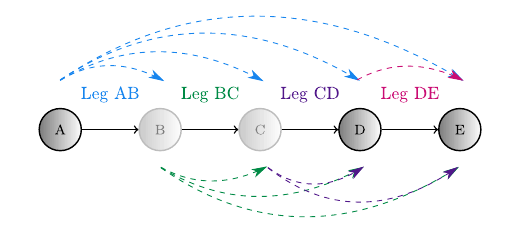}
    \caption{Four-leg-cluster, dotted lines indicate 10 possible itineraries}
    \label{fig:db_4leg}
\end{figure}

The simulation models a network consisting of 5 stations and 4 legs, as shown in Figure \ref{fig:db_4leg}, mirroring the structure of an empirical railway network cutout. The network includes 10 possible itineraries represented by: $\mathcal{O} = \{AB,AC,AD,AE,BC, \allowbreak BD,BE,CD,CE,DE\}$. On each itinerary, the firm offers seven fare classes. In this model, a fare class describes a particular price or fare associated when booking a ticket to travel the itinerary in that class. There are no additional restrictions differentiating classes. 

\subsection{Demand settings}

Extending the demand model described in \citet{Rennie2021} to the network setting, the simulation generates booking requests per customer type $i$ according to a non-homogeneous Poisson process, where the arrival rate per itinerary $o$, \(\lambda_{i,o}(t)\), at time \(t\), is given by:
\begin{equation} \label{eqn:demand1}
    \lambda_{i,o}(t)\vert(D_{o}=d_{o}) = d_{o} \times \phi_{io} \frac{t^{a_{io}-1}(1-t)^{b_{io}-1}}{B(a_{io},b_{io})}.
\end{equation}
Here, $\phi_{io}$ is the fraction of customers of type $i$ and \(D_{o} \sim \mbox{Gamma}(\alpha_{o},\beta_{o})\) with probability density function:
\begin{equation} \label{eqn:demand2}
    f(d_{o}\vert\alpha_{o}, \beta_{o}) = \frac{\beta_{o}^{\alpha_{o}}}{\Gamma(\alpha_{o}) d^{\alpha_{o} -1}e^{\beta_{o} d}},
\end{equation}
where $a_{io}$ and $b_{io}$ define are the parameters of a Beta distribution which defines how customers arrive over time. We generate demand over a horizon of 3,600 time slices to ensure $\lambda_{i,o}(t) < 1$. This level of detail is required to accurately parameterise the dynamic program for bid price control. The resulting bookings are aggregated into 18 booking intervals.

As in \citet{Rennie2021}, we consider differentiated demand from two customer types represented by the set $\mathcal{I} = \{1,2\}$. We assume that customers book the cheapest available fare class and differ in price sensitivity. We define $p_{ijo}$ as the probability that a customer of type $i$ pays up to fare class $j$ on itinerary $o$. By combining demand from two customer types that differ in price sensitivity with offers that depend on the current set of offered classes, we mimic a realistic price effect: Offer prices result from the cheapest class currently offered by the firm, as customers will buy the cheapest available class. When a customer's willingness to pay does not equal or exceed the price of the cheapest available class, they do not buy; hence, their price sensitivity translates to decreased demand. Note that the price-demand response depends on the itinerary and time in the booking horizon.

Combining this demand model with the given network creates 210 demand parameters. Table \ref{tab:params_vals} provides a full list of parameter values, and interpretations of each parameter. We set the parameters to mirror common RM assumptions \citep{Weatherford1992}: (i) valuable customers from type 1 book later than customers from type 2, (ii) customers book earlier for longer journeys, and (iii) customers are willing to pay a higher fare class if they are travelling further. Most passengers book tickets boarding at A and leaving at E; this ensures the correlation between the legs exceeds 0.5 and guarantees that the legs are correctly modelled in the same cluster as detailed in Appendix \ref{app:sim_ver}. 

We validate that the functional dynamical correlation between the four legs for simulated data is comparable to empirical railway data as detailed in Appendix \ref{app:sim_ver}. We generate all regular demand based on these parameters. 

The simulation excludes trend and seasonality to evaluate outlier detection approaches in a best-case scenario. In other words, if an algorithm fails on observations from stationary demand, it will likely not perform better given more demand variability. However, additional results based on simulation data that does feature seasonality can be found in Appendix \ref{app:seasonal}.

\subsection{Outlier generation and evaluation} 
\label{sec:outlier_gen}

We generate demand volume outliers by changing the Gamma distribution parameters that govern the total demand level according to equations~\eqref{eqn:demand1} and \eqref{eqn:demand2}. Previous work found the proportion of outliers had little effect on outlier detection performance in the single-leg case \citep{Rennie2021}. Therefore, we generate booking patterns for 500 departures per demand setting, with 1\% of departures experiencing outlier demand. That is, we generate 495 departures from the regular demand distribution and 5 outliers from a set of twelve outlier distributions where the mean has shifted by $\pm 10\%$, $\pm 20\%$, $\pm 30\%$, $\pm 40\%$, $\pm 50\%$, and $\pm 60\%$. For every shift in the mean, we reduce the variance of the outlier demand distribution by $80\%$. This still results in an overall increase in the variance of total demand in the presence of outliers but also ensures that we sample sufficiently outlying demand values. Outliers may also occur due to factors such as changes in arrival times or changes in customers' willingness to pay. \citet{Rennie2021} provide results on how the performance of functional depth varies under these different types of outliers. Here, we focus on the different types of outliers caused by varying network effects. In all cases, we consider the application of the outlier detection procedure to the constrained demand -- applying the approach directly to the booking patterns without applying any unconstraining approaches first. The problem of unconstraining is one of the major challenges of demand forecasting for revenue management and beyond the scope of this paper.

We differentiate outlier scenarios in terms of the affected network components. Firstly, we evaluate a scenario where outlier demand affects all network itineraries. We consider the case where each outlier is randomly drawn from one of the twelve outlier distributions, resulting in outliers from a mixture of different distributions. This lets us test whether the ranking of the alert list mirrors the outliers' underlying degree of demand deviation. Then, we consider each of the twelve outlier distributions in isolation to assess the detection sensitivity. Secondly, we evaluate a scenario where outliers only affect a single itinerary. This evaluates the benefits of clustering multiple legs. 

In Appendix \ref{app:subset_itin_outliers}, we consider the practically relevant case of outliers affecting a subset of itineraries and provide further details on all simulation experiments. 

Each combination of outcomes can be classified into one of four categories: (i) assigning a non-zero outlier severity to a genuine outlier creates a true positive (TP); (ii) assigning a zero outlier severity to a regular observation creates a true negative (TN); (iii) assigning a non-zero outlier severity to a regular observation creates a false positive (FP); (iv) assigning a zero outlier severity to a genuine outlier creates a false negative (FN). This classification enables us to compute the true positive rate (TPR) for the top $R$ ranked departures in the alert list:
\begin{equation}
    TPR_R = \frac{TP_R}{TP + FN},
\end{equation}
where $TP_R$ is the number of true positives in the top $R$ departures. The true positive rate lies between 0 and 1, where 1 means all genuine outliers were identified. We evaluate performance across 1,000 stochastic simulations.

In an ideal setting, the alert list should feature, from top to bottom, large outliers and, subsequently, smaller outliers. Therefore, we also use the distribution of outliers within the ranked alert list to evaluate how well the method ranks the most critical outliers. 

\subsection{Benchmarked outlier detection approaches} 
\label{sec:bench_approach}

For benchmarking, we term the newly proposed approach FD+Agg and compare it to two alternatives from the literature: Principal Component Analysis combined with High-Density Regions (PCA+HDR) as inspired by \citet{Hyndman2016}, and the leg-based functional depth analysis as proposed in \citet{Rennie2021}.

\subsubsection{Comparison with PCA+HDR} 
This benchmark (i) computes features (e.g. mean, variance, curvature) of the booking patterns for the total demand in a cluster; (ii) uses PCA \citep{Yang2004} to identify the first two principle components from the features; and (iii) uses HDR, a density-based approach \citep{Hyndman1996}, to find the $\nu$ points with the lowest density in the first two principal components. These points are classified as outliers. Extended details of the method, including the list of features, can be found in Appendix \ref{app:benchmark}.  This method provides an ordering of the outliers but not a severity measure, as illustrated by Figure \ref{fig:demand_vol_nh}.

\subsubsection{Comparison with non-ranked, single-leg approaches} 
To highlight to critical features of FD+Agg, we benchmark (i) the use of severity measures to rank outliers and (ii) the inclusion of network effects. To isolate the effects of each of these features, we perform two separate benchmark tests:

\emph{We evaluate the effect of ranking outliers} by measuring the increase in precision when ranking outliers. For example, we consider the precision in the top 5 ranked departures versus 5 randomly chosen departures with non-zero outlier probabilities (i.e., as in \citet{Rennie2021}). The change in precision when considering the top $R$ departures, $\Delta(Precision)_{R}$, is given by:
\begin{equation}
     \Delta(Precision)_{R} = \frac{TP_R}{TP_R + FP_R} - \frac{TP_{R(random)}}{TP_{R(random)} + FP_{R(random)}},
\end{equation}
where $TP_{R(random)}$ is the number of true positives in a random selection of $R$ departures with non-zero severity, and $FP_{R(random)}$ is defined analogously for false positives. 

\emph{We quantify the value of accounting for network effects} by computing ranked alert lists for each leg in isolation. We then compare the true positive rates to the aggregated, network-driven approach presented in this paper. 


\subsection{Detecting outliers in multiple legs} \label{sec:outlier_results}

\begin{figure}[!ht]
    \centering
    \includegraphics[width=\textwidth]{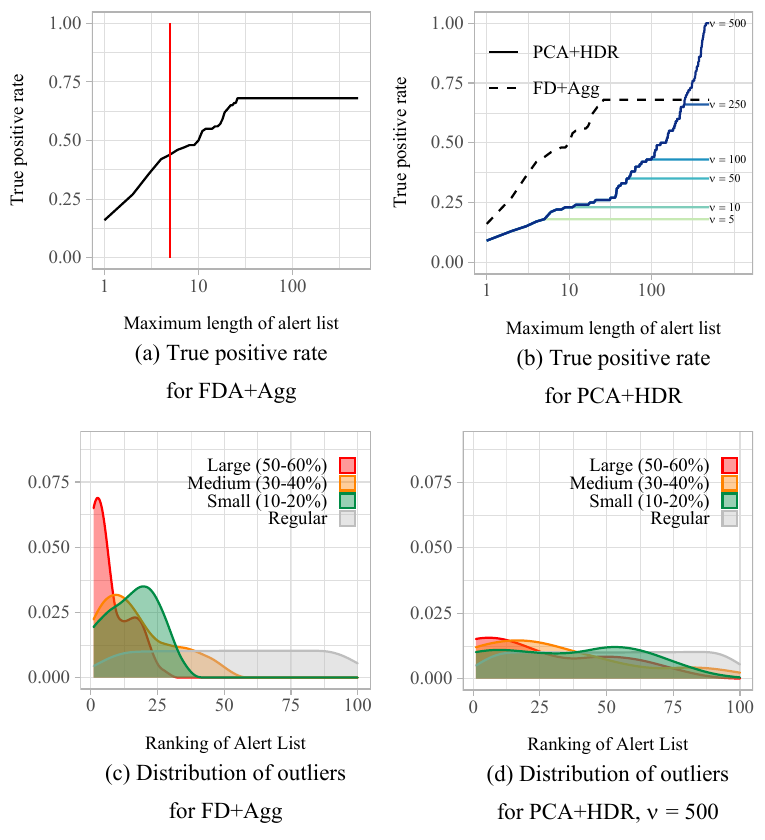}
    \caption{Performance and benchmark comparison with PCA+HDR for demand-volume outliers in all itineraries, showing improved performance}
    \label{fig:demand_vol_nh}
\end{figure}

As a first experiment, we consider the scenario where outlier demand equally affects all itineraries and legs within the cluster. For this scenario, Figure \ref{fig:demand_vol_nh}a illustrates how the true positive rate (TPR) increases when ranking outliers for different lengths of the alert list. The red line indicates the number of genuine outliers. The true positive rates for our method (denoted as FD+Agg) are promising, with a TPR of around 0.2 for a list length of 1. Since there are five genuine outliers, this indicates that a genuine outlier is almost always ranked top. Results under different functional depth thresholds are given in Appendix \ref{app:threshold_results}.

\subsubsection{PCA+HDR benchmark results} The PCA+HDR approach requires a given number of outliers to detect, $\nu$, as input. Therefore, we compare the performance of the benchmark method under different choices of $\nu$ to FD+Agg.

Figure \ref{fig:demand_vol_nh}b shows that the true positive rate achieved by FD+Agg consistently exceeds that achieved by PCA+HDR. To achieve the same level of the true positive rate, PCA+HDR would need to classify around 250 departures (i.e. 50\%) as outliers. In comparison, FD+Agg achieves this rate starting at about 30 classified outliers. We consider this a successful validation of the effect of ranking outliers in FD+Agg. Appendix \ref{app:res_tab} lists these results in tabular format.

Figure \ref{fig:demand_vol_nh}c shows the distribution of each outlier magnitude in the alert lists. Under FD+Agg, the modes of the distributions generally fall where they should, as larger outliers are ranked higher. The smaller variance in the ranking of the larger magnitude outliers indicates that they are easier to detect. The higher variance of the medium-sized outliers can be explained as the ranking of a medium-sized outlier is dependent on which other types of outliers occur: if there is a large and a medium outlier, the medium outlier is ranked lower; if there is a small and a medium outlier, the medium outlier is ranked higher. The distribution of outliers detected by PCA+HDR, shown in Figure \ref{fig:demand_vol_nh}d, also has the modes in the correct order. However, there is much more overlap between the distributions, showing its inability to correctly rank the outliers. 

\subsubsection{Comparison with non-ranked approach}
\begin{figure}[!ht]
    \centering
    \includegraphics[width=\textwidth]{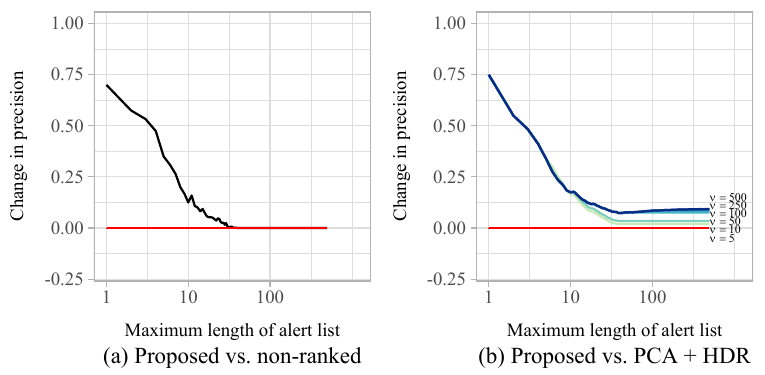}
    \caption{Change in precision from ranking detected outliers in FD+Agg as dependent on the length of the alert list}
    \label{fig:precision_change}
\end{figure}

Figure \ref{fig:precision_change}a highlights how the precision improves when ranking outliers instead of listing them in random order. Ranking particularly improves precision when the alert list covers only a small number of departures. As domain experts indicate that analysts cannot target more than 1\% of departures, ranking focuses resources and thereby provides large benefits in practice. Nevertheless, Figure \ref{fig:precision_change}a (when contrasted with Figure \ref{fig:demand_vol_nh}a) also highlights the trade-off between reducing the number of false alerts and identifying all outliers. A shorter length of alert list increases precision but reduces the true positive rate. 

The increase in precision from applying our method compared to PCA+HDR is similar to the increase in precision from the inclusion of the ranking (see Figure \ref{fig:precision_change}b). This suggests that PCA+HDR performs reasonably well in terms of outlier detection, but poorly in terms of ranking the outliers.

\subsubsection{Comparison with single-leg approach}
\begin{figure}[!ht]
    \centering
    \includegraphics[width=\textwidth]{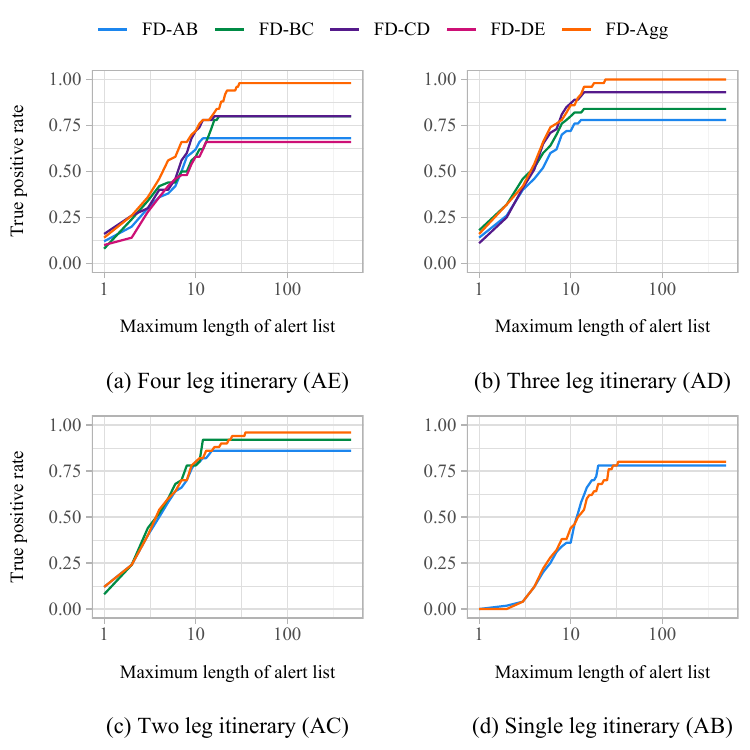}
    \caption{True positive rate for single itinerary outliers when applying FD+Agg versus detection on isolated legs}
    \label{fig:tpr_itin}
\end{figure}

Figure \ref{fig:tpr_itin} shows the true positive rate when a ranked alert list is computed for each leg in isolation versus in the proposed aggregated manner. Here, we consider outlier demand generated by a 50\% increase in the affected legs as an illustrative example. We analyse detection performance by breaking down results in terms of which itinerary the outlier demand is generated in. We show only the results relating to itineraries AB, AC, AD, and AE. Figure \ref{fig:app_tpr_itin} in Appendix \ref{app:single_itin_outliers} details results for the further itineraries yielding similar conclusions. 

For results when outlier demand is generated across combinations of itineraries, refer to Appendix \ref{app:subset_itin_outliers}.

In all cases, the true positive rate for clusters is higher than in any of the individual legs. This is because when considering the leg's bookings in isolation under outlier demand that affects multiple legs, the noise from other itineraries prevents detecting the outlier in every leg. However, clustering increases the number of detected genuine outliers. 

Aggregation is most beneficial when the outlier demand affects the most legs. In our example, this applies when itinerary AE experiences outlier demand, as shown in Figure \ref{fig:tpr_itin}a. The lower true positive rates in legs AB and DE result because different combinations of itineraries also utilise these legs. The aggregation is less beneficial when outlier demand affects an itinerary consisting of only one or two legs since we aggregate the analysis across legs that are actually not affected by outlier demand.  However, there is a modest gain in true positive rate even in this case -- compare Figure \ref{fig:tpr_itin}(c). This is due to the knock-on effects of decreased capacity on the affected legs, impacting the bid prices for any itineraries which include these legs. For some lengths of the alert list, the leg-level true positive rates are higher than the aggregated approach, due to false positives from unaffected legs being included in the list. However, even for itinerary AB (Figure \ref{fig:tpr_itin}d), where false positives from unaffected legs are most likely, the difference is small and cancelled out by the overall increase in true positive rate. 

\subsubsection{Sensitivity to different magnitudes of outliers} 
\begin{figure}[!ht]
    \centering
    \includegraphics[width=\textwidth]{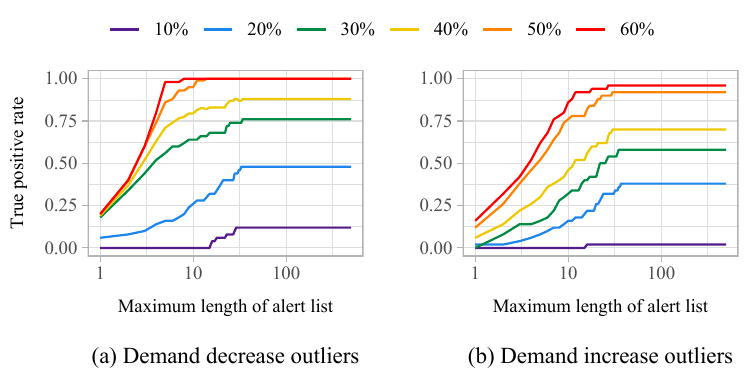}
    \caption{Sensitivity of true positive rate from FD+Agg under different magnitudes of homogeneous demand-volume outliers}
    \label{fig:demand_vol_h_tpr}
\end{figure}

To better understand outlier detection performance, we break down the results by the magnitude of outliers in Figure \ref{fig:demand_vol_h_tpr}. 
When outliers result from minor changes in demand levels, they are difficult to detect, resulting in low true positive rates. Given the significant overlap between the distribution of outlier demand with a 10\% change in magnitude and that of regular demand, this is to be expected. Therefore, 10\% demand changes effectively provide a lower bound on how big an outlier needs to be in order to be detected. 
\begin{figure}[!ht]
    \centering
    \includegraphics[width=\textwidth]{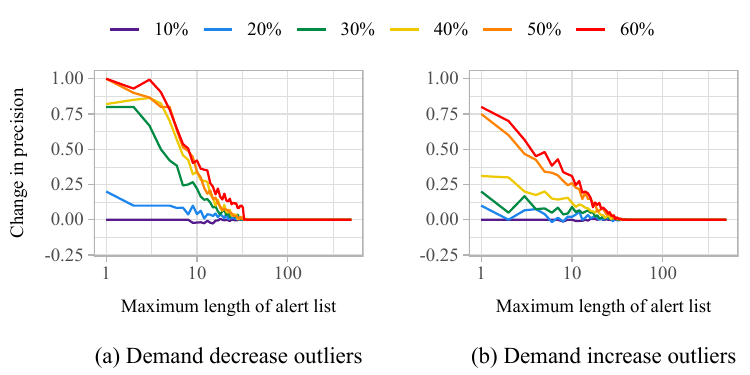}
    \caption{Sensitivity of precision from FD+Agg under different magnitudes of homogeneous demand-volume outliers}
    \label{fig:demand_vol_h_prec}
\end{figure}

As the magnitude of the outliers increases, they become easier to detect and true positive rates are higher, with peak rates reached with shorter alert lists. Thus, genuine outliers are more likely to be ranked higher when they are caused by larger demand changes. For demand decreases of at least 50\%, the true positive rate is very close to the optimal detection rate. Negative demand outliers are slightly easier to detect than positive demand outliers, meaning shorter alert lists are required. This is due to the demand censoring imposed by the booking controls and capacity restrictions. 

Figure \ref{fig:demand_vol_h_prec} shows the precision gap over randomly ordered lists. Once more, larger magnitude outliers result in larger precision improvements from ranking, while detecting minor outliers gains little over random selection. Similarly, we observe that detecting negative demand outliers gains slightly more precision in comparison to detecting positive outliers of the same magnitude. Additional results regarding false discovery rates are available in the appendix.


\section{Simulation Study: Forecast Adjustments} 

To evaluate the implications of adjusting the demand forecast for further planning steps, we simulate network demand and the optimisation of offered fare classes over the booking horizon. We list and explain all parameters determining the settings in the simulation study in Appendix \ref{app:params}. In this section, we first detail how the simulated RM system uses the demand forecast to compute revenue-optimal offers based on bid prices. In that, it follows a widely implemented industry standard. Subsequently, we describe alternative strategies that analysts may apply to adjust demand forecasts based on identified outliers. Finally, by comparing revenue gained from offers based on different adjusted demand forecasts under the same simulated outlier demand, we highlight the effects of adjustments as dependent on outlier scenarios.

\subsection{Network revenue management system}

The simulated RM system controls the offered set of fare classes per itinerary to optimise expected revenue. To that end, it implements a dynamic program to compute bid prices per leg and sums them up per itinerary following the methodology described in \citet{strauss2018review} and detailed in the appendix. To test for the sensitivity of results with regard to the revenue optimisation, we compared two industry standards, the leg-based EMSR heuristic as introduced in \citet{belobaba1987air} and dynamic programming in initial simulations studies not further documented here. The results showed that, for the given demand model, the choice of optimisation approach had little effect on the quality of the outlier detection.

The bid price indicates the marginal difference between the value of selling a seat in the current time period and that of reserving it to sell in a future time period. The RM system only offers fare classes where the revenue from a booking exceeds the bid price. Thus, as an RM term, bid prices do not denote the customer's bid but indicate the minimum price a fare class must carry to be included in the offer set. From those classes in the offer set, customers only consider the cheapest offer.  Bid prices depend on the time until departure, unsold capacity, and expected demand. Note that in the examples given here, we consider a single capacity per leg, not differentiating, for example, 1st or 2nd class compartments with separate capacities.

Booking patterns result as customers arrive and decide to book one of the offered fare classes. The firm does not report booking patterns for each individual itinerary, but only records them on the leg level. 

The dynamic program relies on a given set of expected demand arrival rates per leg $l$, fare class $j$, and time slice $t$ of the booking horizon. In the simulation, we derive expected demand arrival rates from our knowledge of the underlying demand model. Arrival rates for each leg $l$ and fare class $j$ are given as
\begin{equation}
    \hat{\Lambda}_{j,l}(t) = \sum_{o \in \mathcal{O}_l} \sum_{i \in \mathcal{I}} p_{i,j,o} \: \lambda_{i,o}(t),
\end{equation}
where \(\lambda_{i,o}(t)\) is the arrival rate of customers of type $i$ requesting itinerary $o$, and $\mathcal{O}_l$ is the set of itineraries which include leg $l$. This creates an artificially accurate demand forecast. Deriving the demand forecast from the actual demand parameter values ensures that the estimation of revenue loss caused by undetected outliers is not affected by flawed forecasts (see Section \ref{sec:revenue_results}). In practice, demand parameter values are not known but are estimated based on previously observed demand and time series forecasting. A recent survey of related research contributions can be found in \citet{banerjee2020passenger}, while \citet{fiig2019can} represent an example of the ongoing discussion on the link between forecast accuracy and RM performance.

\subsection{Forecast adjustments for outlier demand} \label{sec:sim_adjustment}

One aim of identifying outlier demand in booking patterns is to support analyst adjustments in RM systems. Without such adjustments, offers would be optimised for a regular demand forecast and thereby not be fit for maximising revenue under outlier demand. This raises the difficulty of predicting the consequences of analyst adjustments throughout the network. As a step in this direction, we analyse a best-case scenario, assuming that the adjustment is made with foresight before the start of the booking horizon. We compare the revenue under three different adjustments:
\begin{itemize}
    \item \textbf{Adjustment 1 (conservative)}: Adjust only forecasts of affected single-leg itineraries. E.g. for an outlier creating additional demand for itinerary AC, increase the forecasts of itineraries AB and BC.
    \item \textbf{Adjustment 2 (aggressive)}: Adjust forecasts of all itineraries that include at least one of the affected legs. E.g. for additional demand for itinerary AC, adjust all itineraries, including either leg AB or leg BC -- i.e., itineraries AB, AC, AD, AE, BC, BD, and BE.
    \item \textbf{Adjustment 3 (balanced)}: Adjust forecasts of affected single-leg itineraries and the \textit{cluster-spanning} itinerary -- in this case, AE. E.g. for additional demand for itinerary AC, adjust itineraries AB, BC, and AE. The motivation for adjusting AE (ahead of other itineraries) is that, in general, this will be the most popular itinerary in the cluster.
\end{itemize}
These three adjustments are not the only choices available to analysts. However, they represent options that stretch across the spectrum of how fully network effects should be considered. Adjustments 1 (conservative, leg-based adjustments only) and 2 (aggressive, all potential network effects) are the two extremes. Adjustment 3 (balanced) is a compromise, which is more conservative than Adjustment 2 but still identifies the itinerary most likely to be the source of outlier demand. Further options would be to include more than just the cluster-spanning itinerary in an alternative to Adjustment 3, but this leaves another choice of which itineraries to prioritise. As a lower bound, we compute the revenue when \textbf{no adjustment} is made. As an upper bound, we implement an \textbf{oracle adjustment}, i.e., only adjusting the forecasts of affected itineraries. We compare the revenue as the level of outlier demand ranges from -60\% to +60\% of the average leg demand.

\subsection{Experimental Results: Revenue Benefits} 
\label{sec:revenue_results}

\begin{figure}[!ht]
    \centering
    \includegraphics[width=\textwidth]{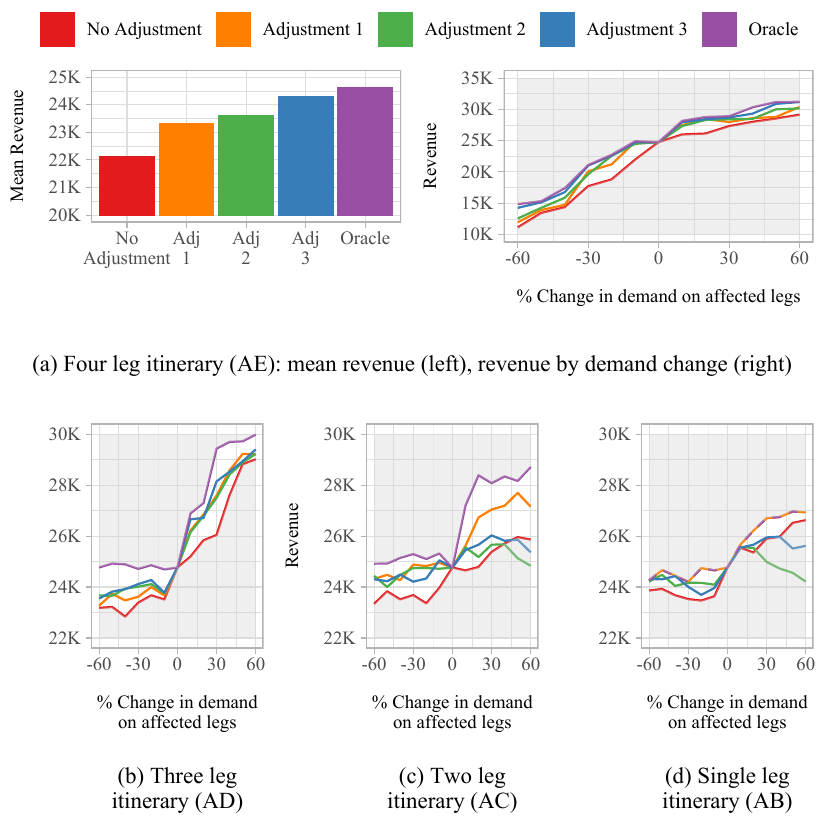}
    \caption{Revenue under under different forecast adjustments; the subtitle indicates the actual outlier source}
    \label{fig:revenue}
\end{figure}

Figure \ref{fig:revenue} shows the revenue generated by outlier demand for each of the three adjustments. We show the results for four of ten itineraries contained within these four legs in Figure \ref{fig:db_4leg}. The results for the other six itineraries are similar. Appendix \ref{app:revenue_results} further details these results as well as results on adjustments after outlier detection.

When outlier demand affects all four legs in the cluster (Figure \ref{fig:revenue}a), any type of adjustment is always better than no adjustment. Besides the oracle, the best choice is Adjustment 3, i.e., the balanced approach, which adjusts the forecasts of the cluster-spanning itinerary and the individual leg. Adjustment 3 is able to obtain, on average, 87\% of the additional revenue gained under the oracle adjustment. Similar results are obtained when the outlier demand affects three legs (Figure \ref{fig:revenue}b). 

When outlier demand affects only a single-leg itinerary (Figure \ref{fig:revenue}d), the conservative Adjustment 1 and the oracle adjustment coincide. The aggressive Adjustment 2 yields less revenue than no adjustment. For example, although leg AB is correctly adjusted, the erroneous adjustment to itineraries AC, AD, and AE results in incorrect forecasts for legs BC, CD, and DE. The asymmetry between adjustment to positive and negative outlier demand is due to the level of demand being bounded from below by 0. Similar results emerge when the outlier affects only two of the affected legs (Figure \ref{fig:revenue}c), though the negative consequences of over-adjusting all potentially affected itineraries are less severe, as this causes fewer superfluous adjustments. 

The negative impact of adjusting unaffected itineraries highlights the importance of correctly clustering legs ahead of outlier detection. The closer the outlier demand itinerary is to the cluster-spanning itinerary, the less risky it is to adjust all affected itineraries within a cluster, and the more benefit can be gained from doing so. From a managerial perspective, the \textit{best} adjustment (other than the oracle) depends on the firm's objective. To maximise revenue when the most common outlier (e.g. itinerary AE) occurs, the balanced Adjustment 3 is preferable. Conversely, if the objective is to minimise risk to revenue even in the more unlikely scenarios (e.g. an outlier in itinerary AB), conservative Adjustment 1 is preferable. Overall, however, there are clear benefits from forecast adjustment.


\section{Empirical Study} 
\label{sec:empirical}

\begin{figure}[!ht]
    \centering
    \includegraphics[width=\textwidth]{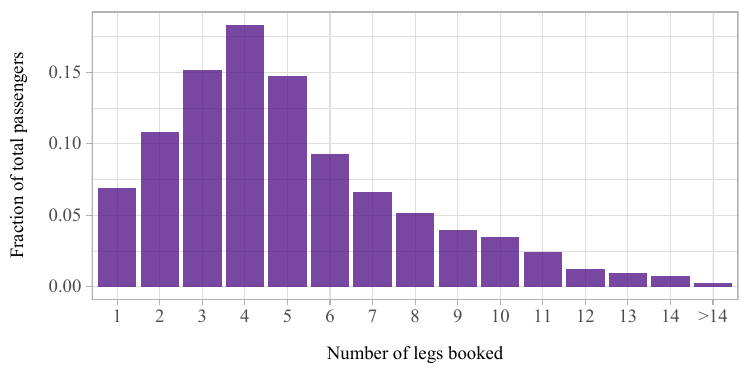}
    \caption{Distribution of the number of legs per booked itinerary from Deutsche Bahn data}
    \label{fig:distribution_leg}
\end{figure}

To demonstrate the practical applicability of the proposed clustering and outlier detection, we apply it to a set of empirical data obtained from Deutsche Bahn. This data set features only bookings of the 2nd class compartment, such that all bookings on one leg require capacity from the same compartment.
The Deutsche Bahn long-distance network consists of over 1,000 train stations, letting the provider offer more than 110,000 direct origin-destination combinations. The numbers grow further when accounting for alternative transfer itineraries and multiple daily departures. Figure \ref{fig:distribution_leg} shows the empirical distribution of the number of legs included in itineraries that passengers booked in November 2019. Only 7\% of passengers booked single-leg itineraries, whereas almost half of all booked itineraries span five or more legs.

\subsection{Clustering legs in the Deutsche Bahn network} \label{sec:db_clustering_comp}

\subsubsection{Small Network Subsection} 
\begin{figure}[!ht]
\centering
\includegraphics[width=\textwidth]{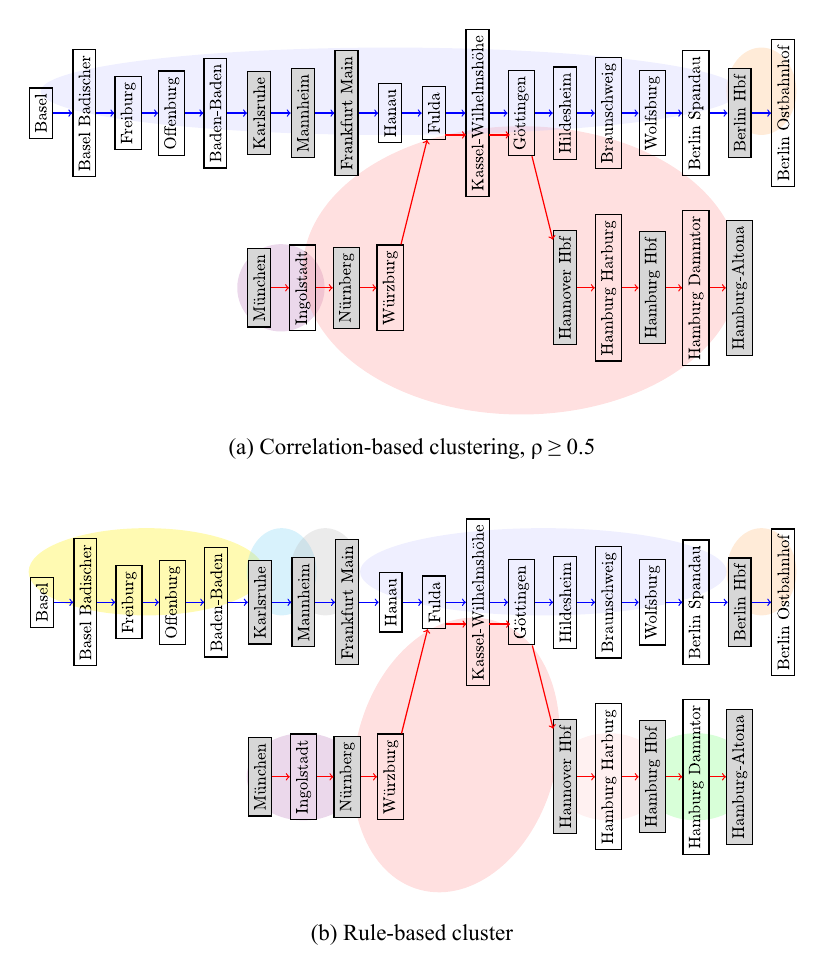}
\caption{Comparison of correlation-based and rule-based clustering of Deutsche Bahn network}
\label{fig:DB_clusters}
\end{figure}

First, we consider a section of the Deutsche Bahn railway network that consists of two intersecting train lines over a total of 27 stations and 28 legs -- see Figure \ref{fig:DB_clusters}. The red train arrives at the connecting stations before the blue train. Hence, the network offers three transfer connections: changing from red to blue at either Fulda, Kassel-Wilhelmsh{\"o}he, or G{\"o}ttingen. This creates 240 potential travel itineraries. For each leg in this network section, Deutsche Bahn records 359 booking patterns for departures between December 2018 and December 2019. Each booking pattern ranges over 19 booking intervals; the first observation occurs 91 days before departure.

We first apply the correlation-based clustering approach of Section~\ref{sec:clustering}, using a threshold of 0.5, such that only legs with a minimum correlation of 0.5 can be in the same cluster. In Figure \ref{fig:DB_clusters}a, coloured bubbles indicate the four resulting clusters: Each train line splits into one large and one small cluster.

To evaluate clustering on empirical data, where the true underlying demand for each itinerary is unknown, we use the network topology to check whether the resulting clusters are plausible. To that end, we propose the following set of rules:
\begin{itemize}
    \item Different train lines must belong to different clusters. Even when passengers can transfer between lines, we expect relatively few passengers to make the same connection. Further, it makes sense to consider train lines separately for forecasting and analyst interventions. 
    \item Train lines are further split into separate clusters on either side of a major station. As many passengers leave the train at a major station and many \textit{different} passengers board, we shall assume a relatively small proportion of passengers book itineraries that pass a major station. Similarly, given that itinerary demand share is driven by which journeys are most common, and passengers often either board or alight at a major station, it is intuitive to have a cluster that contains the legs between major stations.
\end{itemize}
Deutsche Bahn assigns an ordinal indicator of importance to each station, ranging from 1 to 7. We define a \textit{major station} to be in \textit{Category 1}. The entire Deutsche Bahn network includes 21 major stations, whereas the considered network section includes nine major stations. Figure \ref{fig:DB_clusters}b highlights major stations in grey and shows the clusters resulting from the above rules.

The correlation-based clustering returns four clusters, whereas the rule-based clustering returns nine. Nevertheless, the resulting clusters share similar features. Firstly, the two distinct train lines end up in different clusters in either approach. For legs in distinct train lines, correlation tends to be higher between legs that share a transfer station, but not to a convincing extent -- the correlation is at most 0.22. A correlation threshold of 0.27 creates two clusters (one for each train line). Secondly, the breakpoints for the correlation-based approach are a subset of the breakpoints, i.e. major stations, in the rule-based approach. We conclude that the correlation-based approach achieves similar results as the rule-based approach without requiring expert input. 

\begin{figure}[!ht]
    \centering
    \includegraphics[width=\textwidth]{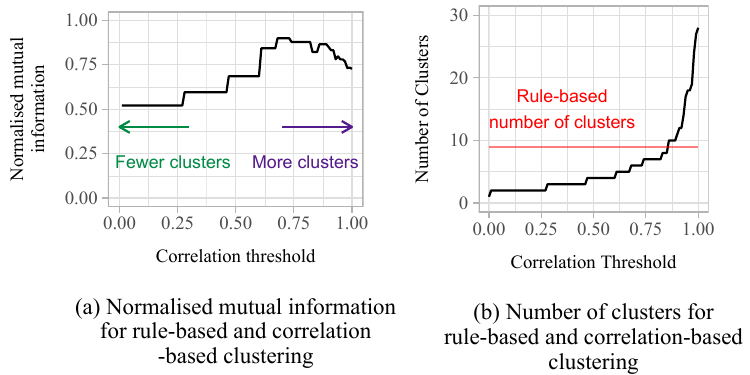}
    \caption{Comparison of rule-based and correlation-based clustering in a two-line railway network}
    \label{fig:db_clust_comp}
\end{figure}

We can formally compare clustering results using the \textbf{Normalised Mutual Information (NMI)} \citep{Amelio2015}. The NMI is 1 if two clusterings are identical, and 0 if they are completely different. 

Figure \ref{fig:db_clust_comp}a shows the NMI between the correlation- and rule-based approaches while varying the threshold in the correlation-based approach from 0 to 1. This shows that both approaches achieve similar results, with an NMI reaching 0.899. The approaches are generally more similar at higher correlation thresholds (around 0.7) since the rule-based approach generally creates more clusters. Figure \ref{fig:db_clust_comp}b compares the number of clusters of the two approaches -- as the correlation threshold changes, the number of clusters ranges from 1 (everything in a single cluster) to 28 (each leg in its own cluster), demonstrating the flexibility of the correlation-based approach.

\subsubsection{Large network subsection} 
\begin{figure}[!ht]
    \centering
    \includegraphics[width=\textwidth]{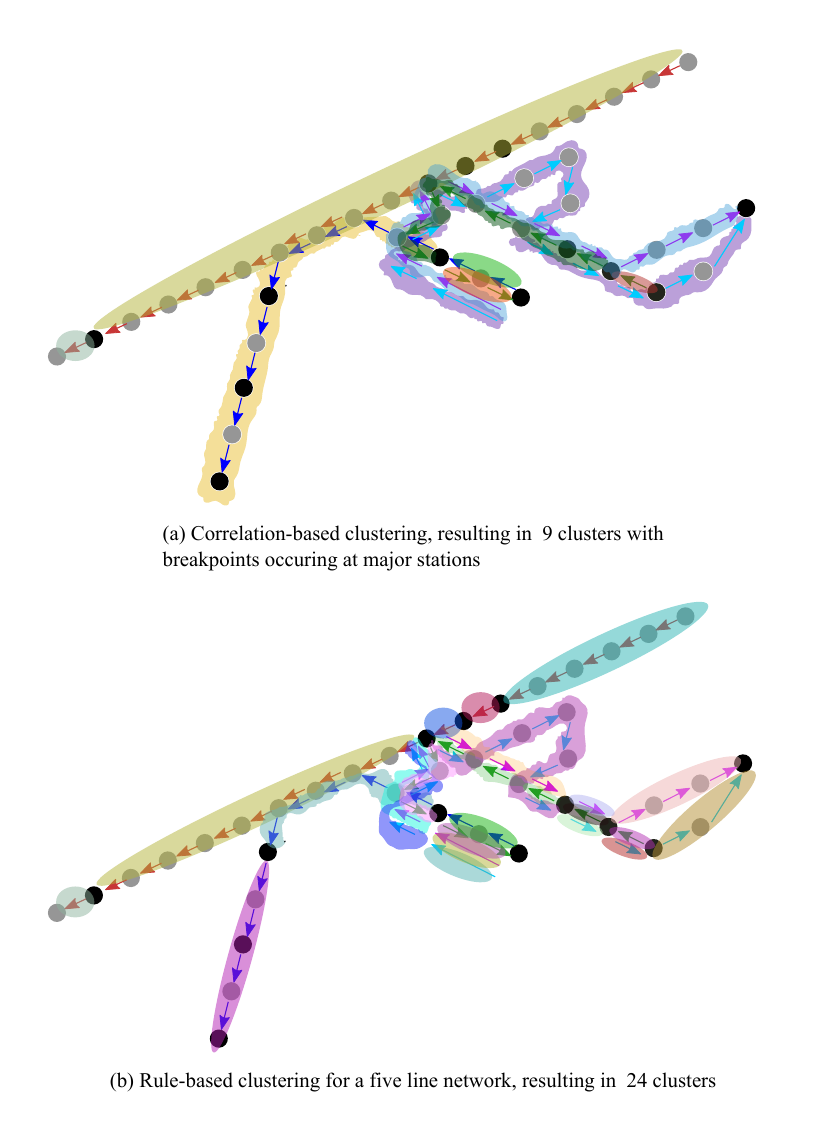}
    \caption{Comparison of rule-based and correlation-based clustering in a five-line railway network}
    \label{fig:db_clust_comp2}
\end{figure}

We extend the empirical study to five train lines to further demonstrate the complexity that considering the network structure brings to clustering and outlier detection, and show the scalability of the approach. The five-line network consists of 40 stations with 63 legs. As seen in Figure \ref{fig:db_clust_comp2}, there are often multiple train lines which cover the same leg or may travel in the opposite direction. As the larger size of the network makes visualisation more difficult, in Figure \ref{fig:db_clust_comp2} stations are represented by circles, with major stations highlighted in black.

Figure \ref{fig:db_clust_comp2}(a) shows the results of the correlation-based clustering with a default threshold  $\rho = 0.5$. This results in 9 clusters, with two train lines each forming their own cluster containing all legs. The breakpoints of the clusters occur at major stations, as also previously seen for two train lines. The pattern of breaking clusters at major stations persists as the correlation threshold is varied. In comparison, the output of the rule-based clustering shown in Figure \ref{fig:db_clust_comp2}(b) results in 24 clusters, with many being of size 1.

In these empirical studies, we applied rule-based clustering only to evaluate the plausibility of the results from correlation-based clustering. We do not advocate for it as a method in itself. A rule-based approach, where the clusters are based on domain experts' categorisations, would not be able to respond to the evolving importance of stations across different train lines and departure times. Notably, the correlation-based method not only uncovers major stations but rather identifies legs where multi-leg itineraries cause similar booking patterns and thus could change and adapt over time. We further evaluate clustering performance in a simulation study, where the itinerary-level demand is known, in Appendix \ref{app:clust_benchmark}. The results in the remainder of the paper rely on correlation-based clustering. 


\subsection{Detecting outliers in the Deutsche Bahn data} 
\label{sec:db_results}

Having established clusters, we apply outlier detection independently to each cluster. To exemplify this on empirical data, we apply the outlier detection procedure to a representative four-leg cluster from the Deutsche Bahn network. Applying the proposed outlier detection approach to empirical data cannot precisely judge detection accuracy, given there is no labelled data on genuine outliers. However, this analysis demonstrates the full process of outlier detection on empirical data including e.g. seasonality and underlines practical implications.

For this analysis, we consider a cluster of four legs from the Deutsche Bahn network with stations anonymised and denoted by A, B, C, D, and E. This cluster results from applying the correlation-based clustering to a new section of the Deutsche Bahn network to Figure~\ref{fig:DB_clusters}. 

\begin{figure}[!ht]
    \centering
    \includegraphics[width=\textwidth]{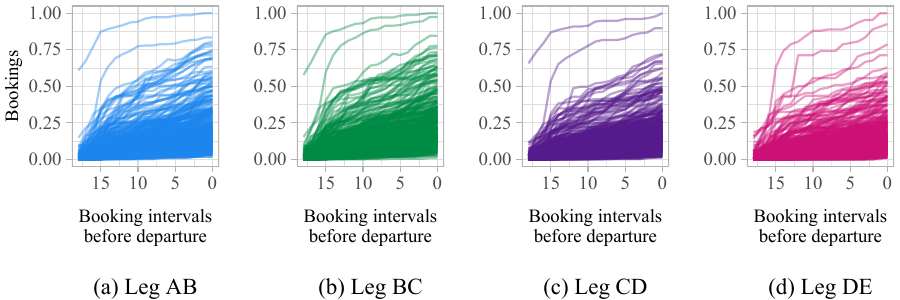}
    \caption{Booking patterns for each leg}
    \label{fig:db_bookings}
\end{figure}

Figure \ref{fig:db_bookings} shows the booking patterns for each of the four legs; bookings are scaled to be between 0 and 1. From initial visual inspection, the structure of the booking patterns appears similar, with some obvious outliers appearing across multiple legs. 

To pre-process the data for outlier detection, we transform the booking patterns by applying a functional regression model \citep{Ramsay1997}. We then apply the outlier detection to the residual booking patterns. In this pre-processing, we correct for three factors: (i) the departure day of the week; (ii) the departure month of the year; and (iii) the length of the booking horizon.\footnote{Deutsche Bahn offers a regular booking horizon of 6 months, with the first observation of bookings occurring around 3 months before departure. Due to schedule changes, shorter booking horizons of 3 months apply for departures from mid-December to mid-March.} 

The functional regression fits a mean function to the booking patterns for each different factor in the model. Table \ref{tab:model_comp_bookings} in Appendix \ref{sec:db_reg} compares models including different factors. Let $y_{nl}(t)$ be the $n^{th}$ booking pattern for leg $l$. Then:
\begin{equation} \label{eqn:db_funcreg}
\begin{split}
  y_{nl}(t) = \beta_{0l}(t) + \textcolor{cyan}{\beta_{1l}(t)\mathds{1}_{Mon_{nl}} + \beta_{2l}(t)\mathds{1}_{Tue_{nl}} + \beta_{3l}(t)\mathds{1}_{Wed_{nl}} +} \\ \underbrace{\textcolor{cyan}{\beta_{4l}(t)\mathds{1}_{Thu_{nl}} + \beta_{5l}(t)\mathds{1}_{Fri_{nl}} + \beta_{6l}(t)\mathds{1}_{Sat_{nl}}+}}_\textrm{\textcolor{cyan}{Departure Day of the Week}} \\ 
  \textcolor{magenta}{\beta_{7l}(t)\mathds{1}_{Jan_{nl}} + \beta_{8l}(t)\mathds{1}_{Feb_{nl}} + \beta_{9l}(t)\mathds{1}_{Mar_{nl}} + } \\ 
  \textcolor{magenta}{\beta_{10l}(t)\mathds{1}_{Apr_{nl}} + \beta_{11l}(t)\mathds{1}_{May_{nl}} + \beta_{12l}(t)\mathds{1}_{Jun_{nl}} + \beta_{13l}(t)\mathds{1}_{Jul_{nl}} + } \\
  \underbrace{\textcolor{magenta}{\beta_{14l}(t)\mathds{1}_{Aug_{nl}} + \beta_{15l}(t)\mathds{1}_{Sep_{nl}} + \beta_{16l}(t)\mathds{1}_{Oct_{nl}} + \beta_{17l}(t)\mathds{1}_{Nov_{nl}} +}}_\textrm{\textcolor{magenta}{Departure Month of the Year}} \\
  \underbrace{\textcolor{blue}{\beta_{18l}(t)\mathds{1}_{Shorter\mbox{ }Horizon_{nl}}}}_\textrm{\textcolor{blue}{Length of Booking Horizon}} + e_{nl}(t).
\end{split}
\end{equation}
where, e.g. $\mathds{1}_{Mon_{nl}} =1$ if departure $n$ relates to a Monday, $0$ otherwise. In this model, $\beta_{0l}(t)$ represents the average bookings for Sunday departures in December, with a regular length of booking horizon, and $\beta_{pl}(t)$ for $p>0$ represent deviations from this mean pattern. The $\beta_{pl}(t)$ are functions of time, which allows for relationships between factors to evolve over the booking horizon. Given that functional depths are calculated independently for each leg, we apply the regression model independently for each leg. The resulting residuals are included in Appendix \ref{app:db_res}, Figure \ref{fig:db_residuals}.
\begin{figure}[!ht]
    \centering
    \includegraphics[width=\textwidth]{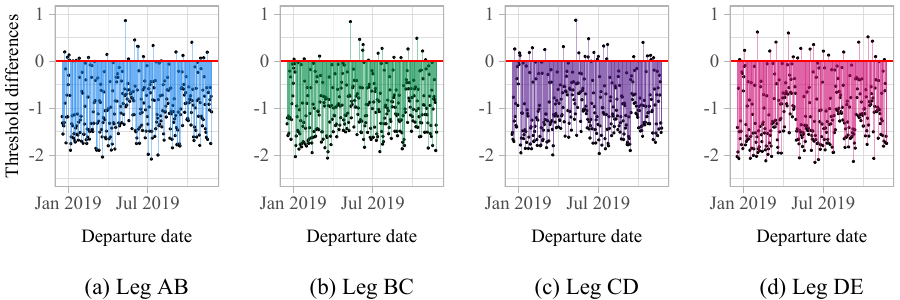}
    \caption{Threshold exceedances per leg, $z_{nl}$}
    \label{fig:db_diffs}
\end{figure}

Functional regression preserves the correlation between different legs, as verified in Appendix \ref{app:sim_ver}, Table \ref{tab:corr_residuals}. The clustering approach can consider either the correlations between the booking patterns or the residual booking patterns. Given that the functional depth (the basis for the outlier detection) is calculated on the residuals, we suggest using the correlation between residual patterns to define the clusters. For this data set, the same clusters resulted in either case. 

We calculate the functional depth of each booking pattern and compute the threshold as described in Section \ref{sec:network_outliers}. We then transform the depths as per equation~\eqref{eqn:z_nl} to obtain $z_{nl}$, as shown in Figure \ref{fig:db_diffs}. The sums of threshold exceedances, $z_n$, were shown earlier in Figure \ref{fig:zn}, with the empirical distribution and fitted generalised Pareto distribution shown in Figures \ref{fig:zn_dist}a and \ref{fig:zn_dist}b, respectively. 

\begin{figure}[!ht]
    \centering
    \includegraphics[width=\textwidth]{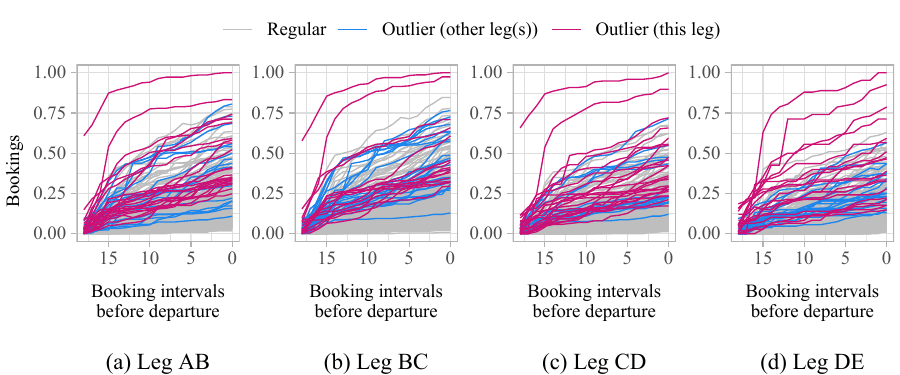}
    \caption{Outliers detected in booking patterns}
    \label{fig:db_outliers}
\end{figure}

Figure \ref{fig:db_outliers} highlights the outliers detected in each leg in pink while depicting outliers detected in other legs but \textit{not} in that leg in blue. Regular patterns are grey. 

Of the 40 outliers (11\% of departures) detected across all legs, 23 outliers (almost 60\%) could be attributed to known events or holidays. When considering only the top 10 outliers, the percentage rose to 70\%. A further departure detected as an outlier had been previously flagged by Deutsche Bahn. The firm implemented a booking stop to control sales on that departure for multiple connected legs. Appendix \ref{app:db_outliers} provides further details on the distribution of identified outliers across legs.


\section{Conclusion and outlook} \label{sec:conclusion}
In this paper, we proposed a two-step method for (i) clustering legs in a mobility network that could benefit from joint outlier detection, and (ii) detecting outlying demand within such clusters. Furthermore, the proposed method, FD+Agg, ranks identified outliers according to their severity, creating an alert list to aid analysts in prioritising demand forecast adjustments. 

The simulation study demonstrated the robustness of the method in a range of outlier demand scenarios. It highlighted that aggregating the analysis across clustered legs improves both detection rate and precision. Further, the ranked alert list often correctly identified the most critical outliers. The advantages of the proposed approach became particularly clear when benchmarking its true positive rate, distribution of outliers across ranks, and precision, against that from a combination of Principal Component Analysis and High-Density Regions (PCA+HDR) from \citet{Hyndman2016}, and on the non-ranked, leg-based method proposed in \citet{Rennie2021}.

Furthermore, we implemented a simulated revenue management system to measure the potential revenue benefits of identifying and adjusting for demand outliers in a network setting by applying forecast adjustments across a cluster of legs. This analysis showed that taking into account the similarity of the legs can improve revenue in most scenarios. In the less likely scenario where only one or two legs of a cluster are affected by outlier demand, risk-averse firms may prefer individual leg-level adjustments. 

Finally, by applying the proposed approach to empirical booking data collected by Deutsche Bahn, we demonstrated its applicability and scalability to the type of data observed in practice. In particular, we used this analysis to showcase the expected cluster results and to demonstrate how to account for additional practical considerations, such as trend and seasonality. Note that once the clustering has been performed, the outlier detection can be performed in parallel within each cluster. Therefore our methodology is scalable to a much larger data set, such as the entire Deutsche Bahn long-distance train network. Such an analysis is not included in this paper as, beyond giving excessive insight into confidential company data, the research insight to be gained from visualising even more complex network cut-outs is limited.

The remainder of this section discusses design choices taken in the research documented here, related limitations, and open research challenges.

\emph{Leg- versus itinerary-level data:} Our proposed method aggregates and analyses booking patterns from \emph{legs} instead of \emph{itineraries} based on three considerations. First, when an extensive network features many possible itineraries, most individual itineraries only receive a small share of bookings, challenging any data analysis -- the study described in Appendix \ref{app:itineraries} evaluates such a case. Though the outlier detection may perform well if there are a sufficient number of bookings for a given itinerary,  only considering such itineraries risks systematically ignoring outliers from smaller itineraries and feeder legs. Secondly, when offering many potential itineraries, providers rarely store all booking patterns per itinerary. For example, capacity-based RM, as described in \citet{strauss2018review}, frequently considers leg booking patterns to ensure capacity availability on each leg of a requested itinerary. Accordingly, the methodology proposed here is compatible with capacity-based RM. Finally, even in the idealised case of having large volumes of stored itinerary-level data for every possible itinerary in the network, then running outlier detection algorithms quickly becomes computationally infeasible as the number of possible itineraries grows rapidly with the size of the network. Detecting outlying clusters of legs, rather than individual itineraries, overcomes all three challenges, as we have demonstrated in this paper. We do however note that the outlier detection methodology we propose could be applied directly to itinerary data without performing clustering. However, we only recommend this for densely booked itineraries, as otherwise, zero-inflated data can induce inferior results. We explored this further in Appendix \ref{app:itineraries}.

\emph{Constrained versus unconstrained bookings:} Observed bookings are constrained by any revenue management controls that were in place at the time of booking, whereas revenue optimisation models rely on unconstrained demand forecasts \cite[Chapter~9.4]{Talluri2004}. To represent this practice, we analysed constrained bookings in this paper and analysed the effect of adjusting unconstrained forecasts in the computational study. In that vein, further research could also consider the impact of applying the analysis to constrained observations, as showcased here, versus applying it to unconstrained demand estimates, which are frequently used for demand forecasting. 

\emph{Implications for decision support:} Further research is needed to consider the practical aspects of outlier detection from the perspective of decision support. Outliers manifest as changes in arrival rate, price elasticity, or other variables that affect bookings.  Outliers can be caused by stochasticity but also by changes in demand patterns as a result of external factors, such as specific events. Complemented by further analysis, successful outlier detection could have three potential uses for RM: 1) Detecting outliers early within the booking horizon through online analysis as proposed in \citet{Rennie2021}, allowing for rapid interventions; 2) removing any detected outliers from training data for demand forecasting to improve results on predicting reference demand curves; and 3) if outliers can be attributed to specific events, the forecast model could be extended to include such events. Outlier detection can have broader benefits for operational planning in transportation networks, helping service providers to avoid overcrowding and delays. To realise such benefits, future research should particularly focus on effective ways to visualise outliers in networks and to communicate alert lists to planners. To further support analysts in their decision-making, additional measures could be included in the alert list. These might include average fare in the affected cluster, potential revenue loss if the outlier is not accounted for, or the outlier severity resulting from running the outlier detection procedure on \emph{revenue} (instead of booking) patterns. An interesting avenue of further research would be to incorporate a feedback element whereby analysts mark outlier alerts as useful or not useful. A supervised learning approach, e.g. one-class-classifiers, could then be combined with our proposed outlier detection routine to filter out false alerts. Analysts could additionally include feedback on the quality of the clustering approach.

\emph{Clustering methodology:} 
Investigating the use of alternative clustering approaches is of interest -- especially where the clusters are likely to be of different structures compared to the rail industry, e.g. in the airline industry where hub and spoke networks are more common than lines. Whilst this paper relied on clustering to improve outlier detection, we believe that the clustering approach is a useful contribution in and of itself. For example, clustering presents additional research avenues such as its application to improving network-level forecasting; supporting the planning for future new stations; evaluating how the transport network structure is changing over time or defining different travel zones. Finally, further research opportunities lie in considering how the success of network outlier detection depends on the network structure. This paper featured examples from transport, specifically railway networks. Other application areas of RM, such as hotels, where correlation is induced by bookings for multiple consecutive nights, feature sparser or structurally different service networks.


\subsection*{Acknowledgements}

We gratefully acknowledge the support of the EP/L015692/1 STOR-i Centre for Doctoral Training funded by the Engineering and Physical Sciences Research Council. The authors thank Deutsche Bahn for the provision of data and are particularly grateful to Philipp Bartke and Valentin Wagner for helpful discussions and suggestions.

\subsection*{Data availability statement}

The simulated data and source code that support the findings of this study are available from the corresponding author upon request.

R functions for outlier detection are available at: \href{https://github.com/nrennie/outlier-detection-in-network-revenue-management}{github.com/nrennie/outlier-detection-in-network-revenue-management}. The code also includes some examples of the simulated data. Functions for simulating demand can be found at \href{https://github.com/nrennie/simnetdemand}{github.com/nrennie/simnetdemand}.

\newpage
\bibliographystyle{apalike}

\newpage
\appendix
\appendixpage
\appendix
\appendixpage

\section{Additional details of method} \label{app:method} 
Appendix \ref{app:method} provides additional details on the proposed method described in Section \ref{sec:method}, including the specifics of the correlation-based minimum spanning tree clustering, and the calculation of the functional depths. 

\subsection{Functional dynamical correlation} \label{app:func_corr}
Let $y_{n,ij}(t)$ be the total observed bookings for the $n^{th}$ departure on leg $ij$ up to booking interval $t$, and similarly for $y_{n,jk}(t)$. The functional dynamical correlation between the booking patterns $y_{n,ij}(t)$ and $y_{n,jk}(t)$ is:
    \begin{equation} \label{eqn:fdc}
        {\rho}_n(ij, jk) = \E \langle y^{*}_{n,ij}(t), y^{*}_{n,jk}(t) \rangle.
    \end{equation}
    where
        \begin{equation}
        \langle y^{*}_{n,ij}(t), y^{*}_{n,jk}(t) \rangle = \int y^{*}_{n,ij}(t) y^{*}_{n,jk}(t) w(t) dt,
    \end{equation}
    and $w(t)$ is a weight function that accounts for the time gap between observations. 
 Here, $y^{*}_{n,ij}(t)$ is a standardised version of $y_{n,ij}(t)$: 
    \begin{equation}
        y^{*}_{n,ij}(t) = \frac{y_{n,ij}(t) - M_{ij} - \mu_{ij}(t)}{\left[ \int\left\{y_{n,ij}(t) - M_{ij} - \mu_{ij}(t)\right\}^2 w(t) dt \right]^{1/2} },
    \end{equation}
where $\mu_{ij}(t)$ is a mean function, and:
\begin{equation}
    M_{ij} = \langle y_{n,ij}(t), 1 \rangle.
\end{equation} 
The functional dynamical correlation is then the average across all $N$ departures:
    \begin{equation}
        \rho(ij, jk) = \frac{1}{N} \sum_{n=1}^{N} \rho_n(ij, jk).
    \end{equation}  

\subsection{Prim's algorithm} \label{app:prims}
Prim's algorithm is a greedy algorithm with the following basic steps. Assuming the original graph $G$ has $V(G)$ vertices.
\begin{itemize}
    \item Initialise the MST, $T$, with the edge with minimum weight and the two vertices it connects. Let $V(T)$ be the number of edges in $T$.
    \item While $V(T) < V(G)$:
    \begin{itemize}
        \item go through the remaining edges in $G$ in order from smallest to largest weights, until one is found that is connected to $T$, but does not form a circuit (i.e. the edge does not form a loop such that $T$ is no longer a tree).
        \item Add this edge (and the vertices it connects) to $T$.
    \end{itemize}
\end{itemize}
More computationally efficient algorithms exist but given the reasonable size of the graphs considered, and more specifically their sparsity (very few stations are adjacent), the computational time is reasonable using Prim's algorithm. 

\subsection{Functional depth} \label{app:func_depth}
The functional halfspace depth is given by:
    \begin{equation}
        d_{nl} (\bm{y}_{nl} \in \mathcal{Y}_l; \alpha) = \sum_{j=1}^{T} w_{\alpha}(t_j) HD_{j}(\bm{y}_{nl}(t_{j})),
    \end{equation}
where, using \(t_{\tau+1} = t_{\tau} + 0.5(t_{\tau} - t_{\tau-1})\), the weights \(w_{\alpha}(t_j)\) are, according to \citet{Hubert2012}: 
    \begin{equation}
        w_{\alpha}(t_j) = \frac{(t_{j+1} - t_j) \mbox{vol}\left[\left\{\bm{x} \in \Real^k : HD_{j}(\bm{x}) \geq \alpha \right\}\right]}{\sum_{j = 1}^{T} (t_{j+1} - t_j) \mbox{vol}\left[\left\{\bm{x} \in \Real^k : HD_{j}(\bm{x}) \geq \alpha \right\}\right]},
    \end{equation}
where $\alpha \in (0,0.5]$, with a default value of $\alpha = 1/{T}$. The sample halfspace depth of a \(K\)-variate vector \(x\) at time \(t_j\) is given by \citep{Hubert2012}:
    \begin{equation}
        HD_{j}(y_{nl}(t_j)) = \frac{1}{N} \min_{\bm{u}, \left\| \bm{u} \right\| = 1} \# \left\{ y_{nl}(t_j), n = 1, \hdots, N: \bm{u}^Ty_{nl}(t_j) \geq \bm{u}^T\bm{x}\right\} 
    \end{equation}

\subsection{Normalised Mutual Information} \label{app:nmi_def}
For a graph containing $M$ legs, the mutual information between two clusterings \(\mathcal{A}\) and \(\mathcal{B}\) of the $M$ nodes in the inverted graph is defined as:
\begin{equation}
    I(\mathcal{A},\mathcal{B}) = \sum_{a=1}^{\lvert \mathcal{A} \rvert} \sum_{b=1} ^{\lvert \mathcal{B} \rvert} \frac{ \lvert \mathcal{A} \cap \mathcal{B} \rvert}{M} \mbox{log} \left( \lvert \mathcal{A} \cap \mathcal{B} \rvert \frac{M}{M_a M_b}\right),
\end{equation}
where $M_a$ is the number of nodes in the $a^{th}$ cluster of clustering $\mathcal{A}$, and similarly for $M_b$. The \textbf{normalised mutual information (NMI)} between two clusterings is defined as \citep{Amelio2015}:
\begin{equation}
    NMI(\mathcal{A},\mathcal{B}) = \frac{2I(\mathcal{A},\mathcal{B})}{H(\mathcal{A})+H(\mathcal{B})},
\end{equation}
where $H(\mathcal{A})$ is the entropy (a measure of uncertainty) defined as:
\begin{equation}
    H(\mathcal{A}) = - \sum_{a=1}^{ \lvert \mathcal{A} \rvert} \frac{M_a}{M} \mbox{log} \left(\frac{M_a}{M}\right).
\end{equation}
$NMI(\mathcal{A},\mathcal{B})$ = 1 if $\mathcal{A}$ and $\mathcal{B}$ are identical, and 0 if they are completely different.

\newpage
\section{Details of computational study} \label{app:comp_study}

Appendix \ref{app:comp_study} contains additional details of the simulation set-up described in Section \ref{sec:simulation}, including the computation of the bid prices, and a validation of the chosen parameter values. 

\subsection{Dynamic programming for bid price control} 
\label{app:dyn_prog}

From \citet{Talluri2004}, let $x$ be the remaining capacity, and define $V_{t}(x)$ denote the value function at time $t$. Define $R(t)$:
\begin{equation}
    R(t) = \begin{cases}
               r_j & \mbox{ if request for fare class $j$ arrives in interval $t$} \\
               0 & \mbox{ otherwise}
            \end{cases}
\end{equation}
where $r_j$ denotes the revenue from accepting a request for fare class $j$. 
The probability that $R(t) = r_j$ is equal to the arrival rate for fare class $j$ at time $t$. Note the arrival rates are such that at most one request arrives in each time period. Define:
\begin{equation}
    u = \begin{cases}
               1 & \mbox{ if request for fare class $j$ arrives \textbf{and} is accepted} \\
               0 & \mbox{ otherwise}
            \end{cases}
\end{equation}
We wish to maximise the combined revenue in the current time period, and the revenue to come in future time periods:
\begin{equation}
    \max_{u \in \{0,1\}} \left(R(t)u + V_{t+1}(x-u) \right)
\end{equation}
The Bellman equation for $V_t(x)$ is:
\begin{eqnarray}
V_t(x) &=& \E \Big[ \max_{u \in \{0,1\}} \left\{R(t)u + V_{t+1}(x-u) \right\} \Big] \\
&=& V_{t+1}(x) + \E \Big[ \max_{u \in \{0,1\}} \left\{(R(t) + \Delta V_{t+1}(x))u \right\} \Big] \\
V_t(x) &=& \sum_{j=1}^{\lvert \mathcal{J} \rvert} \lambda_j(t) max \left\{(r_j - \Delta V_{t+1}(x)),0\right\}
\end{eqnarray}
where $\lambda_j(t)$ is the arrival rate of demand for fare class $j$ in interval $t$:

\begin{equation}
    \lambda_j(t) =  \sum_{i \in I} \lambda_{i,o}(t) p_{ijo},
\end{equation}

and $\Delta V_{t+1}(x) = V_{t+1}(x) - V_{t+1}(x-1)$ is the marginal cost of
capacity in the next time period. The problem is solved with backwards recursion, with the following boundary conditions apply:
\begin{eqnarray}
V_{T+1}(x) = 0,& & \mbox{     }x= 0, 1, \hdots, C \\
V_{t}(0) = 0,& & \mbox{     }t= 1, \hdots, T
\end{eqnarray}
These ensure (i) no revenue can be generated beyond the booking horizon i.e after departure; and (ii) that no further revenue can be generated if there is no capacity remaining. The bid price at time $t$ with remaining capacity $x$ is given by $\Delta V_{t}(x)$.

\subsection{Details of benchmark method} 
\label{app:benchmark}
We use the method proposed by \citet{Hyndman2016} as a benchmark comparison for our proposed method in Section \ref{sec:simulation}. The method works as follows:
\begin{itemize}
    \item Define the total demand booking patterns as the sum of the demand for each leg within the cluster. 
    \item Compute $f$ features of the $n$ total demand booking patterns. Features include: mean, variance, first-order autocorrelation, trend, linearity, seasonality, peak, trough, entropy, lumpiness, spikiness, change in variance, Kullback-Leibler score, among others. See \citet{Hyndman2016} for a full list.
    \item Apply principal component analysis (PCA) as per \citet{Yang2004} to determine the first two principle components i.e. those that explain the most variance.
    \item Use a density-based multi-dimensional approach \citep{Hyndman1996} to find points in the first two principal components with the lowest density.
    \item The $nu$ points with the lowest densities relate to the departures which are classified as outliers. 
\end{itemize}

\newpage
\subsection{Parameter values for simulation study} 
\label{app:params}

The parameter valued used to generate the demand in the computational study are outlined below. 

\begin{longtable}[c]{ c | l | l }
\caption{Regular demand generation parameter values \label{tab:params_vals}}\\
\hline \hline
\multicolumn{1}{c}{\textbf{Parameter}} & \multicolumn{1}{c}{\textbf{Value}} & \multicolumn{1}{c}{\textbf{Effect of parameter}}  \\ \hline
\endfirsthead

\hline
\multicolumn{1}{c}{\textbf{Parameter}} & \multicolumn{1}{c}{\textbf{Value}} & \multicolumn{1}{c}{\textbf{Effect of parameter}} \\ \hline \hline
\endhead

\hline
\endfoot

\hline
\endlastfoot

\begin{tabular}[c]{@{}l@{}}$\bm{\alpha} = \{\alpha_{AB}, \alpha_{AC}, $\\ $\alpha_{AD}, \alpha_{AE}, \alpha_{BC}, $\\ $\alpha_{BD}, \alpha_{BE}, \alpha_{CD}, $\\ $\alpha_{CE}, \alpha_{DE}\}$\end{tabular} &  \begin{tabular}[c]{@{}l@{}}$\bm{\alpha} = \{32, 14, 14,$ \\ $ 180, 4, 4, 14, 4,$ \\ $14, 32\}$\end{tabular}                                                 & \multirow{2}{*}{\begin{tabular}[c]{@{}l@{}} \vspace{-1.4cm} \\ Parameters of the Gamma \\ distribution  which controls the \\ level of total demand across all \\ fare classes and customer types \\ such that the mean demand for \\ itinerary $o$ is: \\ \hspace{2.8cm} $\E(D_{o}) = \frac{\alpha_{o}}{\beta_{o}}$. \end{tabular}}\\ \cline{1-2} 
\begin{tabular}[c]{@{}l@{}}$\bm{\beta} = \{\beta_{AB}, \beta_{AC}, $\\ $\beta_{AD}, \beta_{AE}, \beta_{BC}, $\\ $\beta_{BD}, \beta_{BE}, \beta_{CD}, $\\ $\beta_{CE}, \beta_{DE}\}$\end{tabular}            &  \begin{tabular}[c]{@{}l@{}}$\bm{\beta} = \{1, 1, 1, 1, 1,$\\ $1, 1, 1, 1, 1\}$\end{tabular}    &      \\ \hline
\begin{tabular}[c]{@{}l@{}}$\bm{a_{1}} = \{a_{1,AB}, a_{1,AC}, $\\ $a_{1,AD}, a_{1,AE}, a_{1,BC}, $\\ $a_{1,BD}, a_{1,BE}, a_{1,CD}, $\\ $a_{1,CE}, a_{1,DE}\}$\end{tabular} & \begin{tabular}[c]{@{}l@{}}$\bm{a_{1}} = \{5, 5, 5, 5, 5, $\\ $5, 5, 5, 5, 5\}$\end{tabular} & \multirow{2}{*}{\begin{tabular}[c]{@{}l@{}}\vspace{-1.4cm} \\ Parameters of Beta distribution \\ which controls the arrival \\ times of type 1 customers\end{tabular}}     \\ \cline{1-2}
\begin{tabular}[c]{@{}l@{}}$\bm{b_{1}} = \{b_{1,AB}, b_{1,AC}, $\\ $b_{1,AD}, b_{1,AE}, b_{1,BC}, $\\ $b_{1,BD}, b_{1,BE}, b_{1,CD}, $\\ $b_{1,CE}, b_{1,DE}\}$\end{tabular}   &   \begin{tabular}[c]{@{}l@{}}$\bm{b_{1}} = \{2, 2, 2, 2, 2, $\\ $2, 2, 2, 2, 2\}$\end{tabular}  &   \\ \hline
\begin{tabular}[c]{@{}l@{}}$\bm{a_{2}} = \{a_{2,AB}, a_{2,AC}, $\\ $a_{2,AD}, a_{2,AE}, a_{2,BC}, $\\ $a_{2,BD}, a_{2,BE}, a_{2,CD}, $\\ $a_{2,CE}, a_{2,DE}\}$\end{tabular}  & \begin{tabular}[c]{@{}l@{}}$\bm{a_{2}} = \{2, 2, 2, 2, 2, $\\ $2, 2, 2, 2, 2\}$\end{tabular}  & \multirow{2}{*}{\begin{tabular}[c]{@{}l@{}}\vspace{-1.4cm} \\ Parameters of Beta distribution \\ which controls the arrival \\ times of type 2 customers\end{tabular}}                                   \\ \cline{1-2}
\begin{tabular}[c]{@{}l@{}}$\bm{b_{2}} = \{b_{2,AB}, b_{2,AC}, $\\ $b_{2,AD}, b_{2,AE}, b_{2,BC}, $\\ $b_{2,BD}, b_{2,BE}, b_{2,CD}, $\\ $b_{2,CE}, b_{2,DE}\}$\end{tabular}    &  \begin{tabular}[c]{@{}l@{}}$\bm{b_{2}} = \{2, 3, 5, 7, 2, $\\ $3, 5, 2, 3, 2\}$\end{tabular}  &     \\ \hline
\begin{tabular}[c]{@{}l@{}}$\bm{p_{1jo}} = \{p_{1Ao}, p_{1Oo}$, \\ $p_{1Jo}, p_{1Po}, p_{1Ro},$ \\ $p_{1So}, p_{1Mo}\}$\end{tabular}   &  \begin{tabular}[c]{@{}l@{}}$\bm{p_{1jo}} = \{0.30, 0.25$, \\ $0.20, 0.15, 0.10,$ \\ $0, 0\}$\end{tabular} & \multirow{2}{*}{\begin{tabular}[c]{@{}l@{}}\vspace{-1.1cm} \\ Probability of purchase for each \\ customer type. It is assumed \\ these are constant across \\ itineraries. The no-purchase \\ probability for customer type \\ $i$ is equal to $1 - \sum_{j \in \mathcal{J}}p_{ijo}$.\end{tabular}} \\ \cline{1-2}
\begin{tabular}[c]{@{}l@{}}$\bm{p_{2jo}} = \{p_{2Ao}, p_{2Oo},$ \\ $ p_{2Jo}, p_{2Po}, p_{2Ro},$ \\ $p_{2So}, p_{2Mo}\}$\end{tabular} &  \begin{tabular}[c]{@{}l@{}}$\bm{p_{2jo}} = \{0, 0.05,$ \\ $ 0.10, 0.15, 0.20,$ \\ $0.25, 0.25\}$\end{tabular}  &                            \\ \hline 
$\bm{\phi_o} = \{\phi_{1,o}, \phi_{2,o}\}$ & $\bm{\phi_o} = \{0.5, 0.5\} \forall o$ &
\begin{tabular}[c]{@{}l@{}}\vspace{-0.6cm} \\ Proportion of total demand \\ from 
 each customer type for \\ each itinerary. It is assumed \\ these are constant across \\ itineraries.\end{tabular}  \\ \hline
\begin{tabular}[c]{@{}l@{}}$\bm{x} = \{x_{AB}, x_{BC},$\\$x_{CD}, x_{DE}\}$\end{tabular} & \begin{tabular}[c]{@{}l@{}}$\bm{x} = \{200, 200,$\\$200, 200\}$\end{tabular}  &
\begin{tabular}[c]{@{}l@{}}\vspace{-0.6cm} \\ Total capacity of leg\end{tabular}  \\ \hline
$A, O, J, P, R, S, M$ & \begin{tabular}[c]{@{}l@{}}
1 leg:$\{70,60,50,40,$\\$30,20,10\}$ \\
2 legs: $\{126,108,90,$\\$72,54,36,18\}$ \\
3 legs:$\{168,144,120,$\\$96,72,48,24\}$ \\
4 legs:$\{196,168,140,$\\$112,84,56,28\}$ \\
\end{tabular} &
\begin{tabular}[c]{@{}l@{}}\vspace{-0.6cm} \\ Set of fare classes and \\ associated prices. \end{tabular}  \\ \hline \hline       
\end{longtable}

\subsubsection{Outliers considered in computational study}
\label{app:outlier_params}

Table \ref{tab:sim_study} shows the different experiments that were carried out as part of the computational study. We consider \textit{cluster} outliers in which every itinerary within the cluster is equally affected; \textit{itinerary} outliers where only a single itinerary within the cluster is affected; and \textit{station} outliers which affect all itineraries that end at a particular station. 

\begin{table}[!h]
\centering
\begin{tabular}{cccc}
\hline \hline
\textbf{Experiment} & \textbf{Outlier Type}                & \textbf{Itineraries Affected}      & \textbf{Magnitudes}    \\ \hline
1 & \textbf{Cluster} & All                                & \begin{tabular}[c]{@{}l@{}}+10\%, +20\%, +30\%, +40\%,\\ +50\%, +60\%, -10\%, -20\%,\\ -30\%, -40\%, -50\%, -60\%\end{tabular} \\ \hline
2 & \multirow{10}{*}{\textbf{Itinerary}} & AB  & +50\%  \\
3 &                                     & AC  & +50\%  \\
4 &                                     & AD  & +50\%  \\
5 &                                     & AE  & +50\%  \\
6 &                                     & BC  & +50\%  \\
7 &                                     & BD  & +50\%  \\
8 &                                     & BE  & +50\%  \\
9 &                                     & CD  & +50\%  \\
10 &                                     & CE  & +50\%  \\
11 &                                     & DE  & +50\%  \\ \hline
12 & \multirow{4}{*}{\textbf{Station}} & AB  & +50\%  \\
13 &  & AC, BC  & +50\% \\
14 &  & AD, BD, CD  & +50\%    \\
15 &  & AE, BE, CE, DE & +50\%  \\ \hline \hline
\end{tabular}
\caption{Different types of outliers considered in computational study}
\label{tab:sim_study}
\end{table}

\newpage
\section{Computational results} \label{app:results}
Appendix \ref{app:results} includes the extended results from the computational study described in Section \ref{sec:simulation}. Results from additional simulation experiments to test the proposed clustering approach are also presented here. 

\subsection{Evaluation of network clustering} \label{app:clust_benchmark}
\begin{figure}[!ht]
    \centering
    \includegraphics[width=\textwidth]{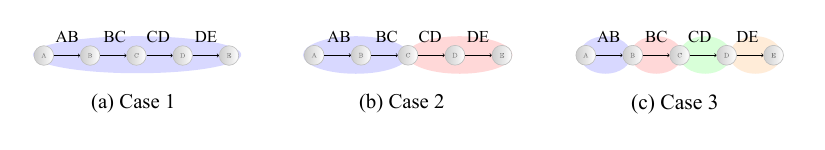}
    \caption{Benchmark clustering}
    \label{fig:benchmark}
\end{figure}

For the correlation-based clustering to perform well it needs to (i) accurately estimate similarity between adjacent legs, and (ii) use information about the pairwise similarity between adjacent legs to detect similarity between (potentially) more than two legs to form clusters. We use the proportion of total demand belonging to each itinerary to determine a clustering benchmark. For example, in Figure \ref{fig:benchmark}a, when \textit{all} passengers travel the itinerary from A to E, the resulting bookings in each of the four legs would be identical. In this case, the correlation between legs would be 1 -- giving a single cluster of four legs. 

To evaluate the clustering when the underlying demand is known, we define the \textbf{common traffic ratio} between two adjacent legs as the proportion of total demand that relates to itineraries over both legs. That is, for two legs $ij$ and $jk$, we define the common traffic ratio, $r(ij,jk)$, to be: 
\begin{equation}
    r(ij,jk) = \frac{D_{ik}}{D_{ij}+D_{jk}+D_{ik}},
\end{equation}
where $D_{ij}$ is the demand for itinerary $ij$, and $D_{ik}$ is the total demand for all itineraries which include both legs $ij$ and $jk$. If all passengers book itineraries that traverse both legs, then $r(ij,jk) = 1$. Conversely, if no passengers book journeys that traverse both legs, then $r(ij,jk) = 0$. 

We vary the level of demand for each itinerary to generate different benchmark clusterings. The output of the correlation-based clustering is then compared with benchmark clustering using the NMI. We consider three cases: the four legs belong in a single cluster (Figure \ref{fig:benchmark}a); they belong in two clusters (Figure \ref{fig:benchmark}b); and they belong in four clusters (Figure \ref{fig:benchmark}c). 

\begin{figure}[!ht]
    \centering
    \includegraphics[width=\textwidth]{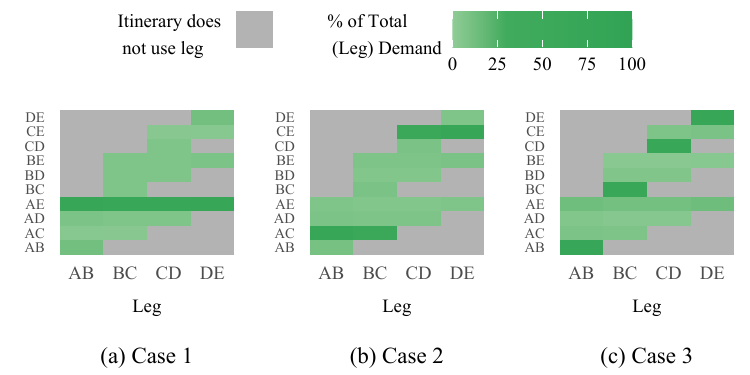}
    \caption{Itinerary demand per leg}
    \label{fig:demand_itin}
\end{figure}

\begin{itemize} 
    \item \textbf{Case 1}: When itinerary AE accounts for at least 50\% of the network demand, we expect legs AB, BC, CD, and DE to belong to the same cluster, as they experience mostly the same demand. The remaining demand is calibrated across itineraries such that the total demand for each leg is reasonably uniformly distributed. We compare the correlation-based clustering with the benchmark clustering of all four legs in a single cluster, when the average percentage of demand on each leg from itinerary AE is 50\%, 60\%, 70\%, 80\%, 90\%, or 100\%. Figure \ref{fig:demand_itin}a shows the fraction of total demand on each leg, from each itinerary, in the case where 60\% of demand is for itinerary AE.
    \item \textbf{Case 2}: We calibrate the majority of demand on leg AB and BC to be for itinerary AC, and the majority of demand on legs CD and DE to be demand for itinerary CE. For simplicity, the distribution of demand is symmetric across the four legs. We compare the performance when the average percentage of demand on each leg belonging to the clustering benchmark itinerary is 50\%, 60\%, 70\%, 80\%, 90\%, or 100\%. Figure \ref{fig:demand_itin}b shows the case where 60\% of demand on each leg is for the respective cluster itineraries (AC or CE). 
    \item \textbf{Case 3}: We calibrate the majority of demand on leg AB for itinerary AB, the majority of demand on leg BC for itinerary BC, and so on. We compare the performance when the average percentage of demand on each leg belonging to the leg itinerary is 50\%, 60\%, 70\%, 80\%, 90\%, or 100\%. Figure \ref{fig:demand_itin}c shows the case where 60\% of demand on each leg is for the itinerary consisting of only that leg.
\end{itemize}
The results are shown in Table \ref{tab:nmi_clustering}. 

\begin{table}[!ht]
\centering
\resizebox{\textwidth}{!}{\begin{tabular}{c|>{\centering}p{1.75cm} >{\centering}p{1.75cm} >{\centering}p{1.75cm} >{\centering}p{1.75cm} >{\centering}p{1.75cm} p{1.75cm}}
\hline \hline 
\multirow{2}{*}{} & \multicolumn{6}{c}{\textbf{Fraction of Leg Demand Resulting from Cluster Itinerary Demand}} \\ \cline{2-7}
& \multicolumn{1}{c}{\textbf{50\%}}    & \multicolumn{1}{c}{\textbf{60\%}} & \multicolumn{1}{c}{\textbf{70\%}} & \multicolumn{1}{c}{\textbf{80\%}} & \multicolumn{1}{c}{\textbf{90\%}} & \multicolumn{1}{c}{\textbf{100\%}} \\ \hline
\textbf{Case 1} & 0.99 & 1.00 & 1.00  & 1.00 & 1.00 & \multicolumn{1}{c}{1.00} \\ 
\textbf{Case 2} & 0.98 & 0.99 & 1.00  & 1.00 & 1.00 & \multicolumn{1}{c}{1.00}  \\ 
\textbf{Case 3} & 0.94 & 0.97 & 0.99  & 1.00 & 1.00 & \multicolumn{1}{c}{1.00} \\ 
\hline \hline \end{tabular}}
\caption{Normalised mutual information}
\label{tab:nmi_clustering}
\end{table}

In almost all cases, the normalised mutual information between the correlation-based clustering and the benchmark equals 1, indicating congruence. We now extend the simulation study by comparing the output of the correlation-based clustering under different correlation measures. In additional to the functional dynamical correlation measure described in Section \ref{sec:clustering}, we compare \textit{Pearson correlation} \citep{Pearson1895} and \textit{Kendall rank correlation} \citep{Kendall1938}. Let $y_{n,ij}(t)$ be the observed bookings for the $n^{th}$ departure on leg $ij$, and $y_{n,pq}(t)$ analogous for leg $pq$.
\begin{itemize}
    \item \textbf{Pearson correlation}: calculate the Pearson correlation between corresponding booking patterns, then average across all booking patterns. That is, for the $n^{th}$ of $N$ booking patterns observed over $T$ booking intervals, we calculate the Pearson correlation coefficient as:
    \begin{equation} 
        \rho_n(ij, pq) = \frac{\sum_{t=1}^{T}(y_{n,ij}(t) - \widebar{y_{n,ij}})(y_{n,pq}(t) - \widebar{y_{n,pq}})}{\sqrt{\sum_{t=1}^{T}(y_{n,ij}(t) - \widebar{y_{n,ij}})^2}\sqrt{\sum_{t=1}^{T}(y_{n,pq}(t) - \widebar{y_{n,pq}})^2}}
    \end{equation}
    where $\widebar{y_{n,ij}}$ is the mean number of bookings for the $n^{th}$ booking pattern. Then:
    \begin{equation}
        \rho(ij, pq) = \frac{1}{n}\sum_{n=1}^{N} \rho_n(ij, pq).
    \end{equation}
    \item \textbf{Kendall rank correlation}: observations $(y_{n,ij}(s), y_{n,pq}(s))$ and $(y_{n,ij}(t), y_{n,pq}(t))$ where $s < t$, are \textit{concordant} if their ordering agrees, and \textit{discordant} otherwise. The Kendall rank correlation is defined between the $n^{th}$ booking patterns in legs $ij$ and $pq$ as:
    \begin{equation} 
        \rho_n(ij, pq) = \frac{t_c - t_d}{\sqrt(t_0 - t_1)(t_0-t_2)}
    \end{equation}
    where $t_c$ is the number of concordant pairs, $t_d$ is the number of discordant pairs, and $t_0$, $t_1$, and $t_2$ are defined as follows:
    \begin{equation}
        t_0 = \frac{T(T-1)}{2},
    \end{equation}
    \begin{equation}
        t_1 = \sum_{s} u_s(u_s - 1) /2,
    \end{equation}
    \begin{equation}
        t_2 = \sum_{t} v_t(v_t - 1) /2,
    \end{equation}
    where $u_s$ is the number of tied values in the $s^{th}$ group of ties for in booking patterns for leg $ij$, and $v_t$ is analogous for leg $pq$. Then:
    \begin{equation}
        \rho(ij, pq) = \frac{1}{n}\sum_{n=1}^{N} \rho_n(ij, pq).
    \end{equation}
\end{itemize}
We compare the cases where the correlation measure is (i) applied directly to the booking patterns, and (ii) applied to the booking patterns where the within-booking pattern relationships e.g. trends have been removed. The normalised mutual information between the clustering produced by the correlation-based clustering under each of the different correlation measures and the benchmark clustering is shown in Table \ref{tab:nmi_app}. 

\begin{table}[!ht]
\centering
\resizebox{1\textwidth}{!}{\begin{tabular}{c|l|p{1cm}p{1cm}p{1cm}p{1cm}p{1cm}p{1cm}}
\hline \hline
\multicolumn{1}{c}{\multirow{2}{*}{\textbf{Case}}} & \multicolumn{1}{c}{\multirow{2}{*}{\textbf{Correlation Measure}}} & \multicolumn{6}{c}{\textbf{\thead{Fraction of Leg Demand Resulting \\ from Cluster Itinerary Demand}}}                           \\
\multicolumn{1}{c}{}  & \multicolumn{1}{c}{}                                             & \textbf{50\%} & \textbf{60\%} & \textbf{70\%} & \textbf{80\%} & \textbf{90\%} & \textbf{100\%} \\ \hline
{\multirow{8}{*}{Case 1}} & \multicolumn{7}{c}{Booking patterns}    \\ \cline{2-8}
& Functional dynamical correlation   &  0.99  &  1.00  &  1.00  & 1.00  & 1.00  &  1.00  \\
& Pearson correlation                                               &              1.00 &   1.00   &  1.00  &  1.00  &  1.00  &  1.00 \\
& Kendall rank correlation                                          &              1.00 &   1.00   &  1.00  &  1.00  &  1.00  &  1.00 \\        \cline{2-8}
& \multicolumn{7}{c}{Differenced booking patterns}  \\ \cline{2-8} 
& Functional dynamical correlation   &  0.99  &  1.00  &  1.00  & 1.00  & 1.00  &  1.00  \\
& Pearson correlation                                               &              0.98 &   1.00   &  1.00  &  1.00  &  1.00  &  1.00 \\
& Kendall rank correlation                                          &              1.00 &   1.00   &  1.00  &  1.00  &  1.00  &  1.00 \\   \hline 
{\multirow{8}{*}{Case 2}} & \multicolumn{7}{c}{Booking patterns}    \\ \cline{2-8}
& Functional dynamical correlation   & 0.98 & 0.99 & 1.00  & 1.00 & 1.00 & 1.00   \\
& Pearson correlation     &   0.00  & 0.00 & 0.00 & 0.00  & 0.00   & 0.00 \\
& Kendall rank correlation &   0.00  & 0.00 & 0.00 & 0.00 & 0.00 & 0.00 \\        \cline{2-8}
& \multicolumn{7}{c}{Differenced booking patterns}  \\ \cline{2-8}
& Functional dynamical correlation   & 0.98 & 0.99 & 1.00  & 1.00 & 1.00 & 1.00   \\
& Pearson correlation      &   0.00  & 0.00 & 0.00 & 0.00 & 0.00 & 0.00 \\
& Kendall rank correlation         &   0.00  & 0.00 & 0.00 & 0.00 & 0.00 & 0.00 \\ \hline
{\multirow{8}{*}{Case 3}} & \multicolumn{7}{c}{Booking patterns}    \\ \cline{2-8}
& Functional dynamical correlation   & 0.94 & 0.97 & 0.99  & 1.00 & 1.00 & 1.00   \\
& Pearson correlation     &   0.00  & 0.00 & 0.00 & 0.00  & 0.00   & 0.00 \\
& Kendall rank correlation &   0.00  & 0.00 & 0.00 & 0.00 & 0.00 & 0.00 \\        \cline{2-8}
& \multicolumn{7}{c}{Differenced booking patterns}  \\ \cline{2-8}
& Functional dynamical correlation   & 0.93 & 0.96 & 0.99  & 1.00 & 1.00 & 1.00   \\
& Pearson correlation      &   0.00  & 0.00 & 0.00 & 0.00 & 0.00 & 0.00 \\
& Kendall rank correlation         &   0.00  & 0.00 & 0.00 & 0.00 & 0.00 & 0.00 \\
\hline \hline  
\end{tabular}}
\caption{Normalised mutual information under different correlation measures}
\label{tab:nmi_app}
\end{table}

For case 1, all three correlation measures seem to be performing equally well, with the normalised mutual information almost always indicating congruence. For cases 2 and 3, the Pearson and Kendall correlation results in extremely poor performance in terms of NMI, with the benchmark clustering never being achieved. Functional dynamical correlation, however, continues to perform well with an NMI close to 1. 

\begin{figure}[!ht]
\centering
    \includegraphics[width=0.5\textwidth]{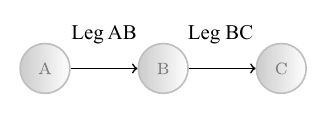}
\caption{Network with two legs}
\label{fig:2leg_app}
\end{figure}

In order to determine why the Pearson and Kendall rank correlations initially appear to perform well in the single cluster case but fail in the two cluster case, we also compare the value of the correlation coefficient with the known demand share in a simple two-leg example. Consider the simple two-leg network shown in Figure \ref{fig:2leg_app}.

The common traffic ratio of legs AB and BC is:
\begin{equation}
    r(AB,BC) = \frac{D_{AC}}{D_{AB}+D_{BC}+D_{AC}},
\end{equation}
If $r(AB,BC) =1$, then the number of bookings on leg AB and leg BC are identical, and the correlation between them is 1. Conversely, if $r(AB,BC) =0$, then the bookings on leg AB and leg BC are independent with correlation 0. Table \ref{tab:corr_comp} shows the estimates of the correlation, compared to the true ratio, $r(AB,BC)$.  

\begin{table}[!h]
\resizebox{\textwidth}{!}{\begin{tabular}{l|ccccccccccc}
\hline \hline \textbf{$r(AB,BC)$} & \textbf{0.0}          & \textbf{0.1}  & \textbf{0.2} & \textbf{0.3}  & \textbf{0.4}  & \textbf{0.5}  & \textbf{0.6}  & \textbf{0.7} & \textbf{0.8}  & \textbf{0.9}  & \textbf{1.0}     \\ \hline 
\multicolumn{12}{c}{Correlation between booking patterns}    \\ \hline
Functional dynamical correlation & 0.12                   & 0.22 & 0.35 & 0.40 & 0.46 & 0.55  & 0.66 & 0.82 & 0.86 & 0.90 & 1.00 \\ 
Pearson correlation   & 0.99  & 0.99  & 0.99  & 1.00   & 1.00        & 1.00  & 1.00  & 1.00  & 1.00  & 1.00  & 1.00 \\ 
Kendall rank correlation  & 0.99  & 0.99  & 0.99  & 0.99  & 0.99 & 0.99  & 0.99  & 0.99  & 1.00  & 1.00  & 1.00 \\ \hline
\multicolumn{12}{c}{Correlation between differenced booking patterns}    \\ \hline
Functional dynamical correlation & 0.14 & 0.18 & 0.29 & 0.42  & 0.50  & 0.53 & 0.66 & 0.83  & 0.88 & 0.91 & 1.00 \\ 
Pearson correlation   & 0.70 & 0.71  & 0.77  & 0.82  & 0.85  & 0.89  & 0.92 & 0.95 & 0.96  & 0.98  & 1.00 \\
Kendall rank correlation & 0.88 & 0.90  & 0.91 &  0.91 & 0.92 & 0.94  & 0.94 & 0.95 & 0.96 & 0.97 & 1.00 \\ \hline \hline
\end{tabular}}
\caption{Comparison of correlation measures}
\label{tab:corr_comp}
\end{table}

Functional dynamical correlation applied directly to the data performs best in all cases. In case 1, where the benchmark clustering is a single cluster, poor clustering performance can only result from underestimating the demand share. Both Pearson and Kendall rank correlation over-estimate the correlation between booking patterns, even when the within-booking pattern effects have been removed. This explains the good performance of Pearson and Kendall rank correlation in case 1, despite extremely poor performance in cases 2 and 3. 

\subsection{Detecting outliers in multiple legs} 
\label{app:outlier_results}

\subsubsection{Results on detecting outliers in multiple legs} \label{app:res_tab}

Table \ref{tab:main_results} provides the results shown in Figure \ref{fig:demand_vol_nh} in tabular format.

\begin{table}[!h]
\centering
\begin{tabular}{lccccccc}
\hline \hline
\textbf{Length of alert list} & \textbf{1} & \textbf{5} & \textbf{10} & \textbf{50} & \textbf{100} & \textbf{250} & \textbf{500} \\ \hline
FD+Agg                      & 0.16       & 0.44       & 0.50        & 0.68        & 0.68         & 0.68         & 0.68         \\
PCA+HDR ($\nu = 5$)         & 0.09       & 0.18       & 0.18        & 0.18        & 0.18         & 0.18         & 0.18         \\
PCA+HDR ($\nu = 10$)        & 0.09       & 0.18       & 0.23        & 0.23        & 0.23         & 0.23         & 0.23         \\
PCA+HDR ($\nu = 50$)        & 0.09       & 0.18       & 0.23        & 0.35        & 0.35         & 0.35         & 0.35         \\
PCA+HDR ($\nu = 100$)       & 0.09       & 0.18       & 0.23        & 0.35        & 0.43         & 0.43         & 0.43         \\
PCA+HDR ($\nu = 250$)       & 0.09       & 0.18       & 0.23        & 0.35        & 0.43         & 0.66         & 0.66         \\
PCA+HDR ($\nu = 500$)       & 0.09       & 0.18       & 0.23        & 0.35        & 0.43         & 0.66         & 1   \\ \hline \hline        
\end{tabular}
\caption{True positive rate of FD+Agg in comparison to PCA+HDR benchmark under varying lengths of alert list}
\label{tab:main_results}
\end{table}

\subsubsection{Outlier detection under alternative thresholds} \label{app:threshold_results}

We recognise that the percentage of departures that analysts are able to adjust strongly depends on the ratio of analysts to departures and that this is likely to be domain-dependent. Therefore, here we consider outlier detection performance as the functional depth threshold varies.

In terms of true positive rates, the threshold choices of 0.01, 0.05, or 0.1 produce similar results, at least near the top of the alert list. Our method ranks the departures classified as outliers such that genuine outliers are more likely to be at the top of the ranked list, and false positives at the bottom of the list. Therefore, using a higher threshold tends to add more departures to the bottom of the list, and increase the risk of more false positives. As shown in Section 4 of the manuscript, the outliers that a threshold of 0.01 fails to detect tend to be small changes in magnitude. It is these small magnitude outliers that are added to the bottom of the list as the threshold increases. Notably, a threshold of 0.001 results in reduced performance even at the top of the list, suggesting this would be too low a threshold. Similar results are seen in the change in precision (compared to the non-ranked method with the same threshold).

A higher threshold does result in higher overall true positive rates as more departures are classified as outliers. However, the maximum true positive rate for a threshold of 0.05 results in around 1 in 5 departures being classified as outliers. This is quite a high percentage for them all to be considered \textit{outliers}.
\begin{figure}[!ht]
    \centering
    \includegraphics[width=\textwidth]{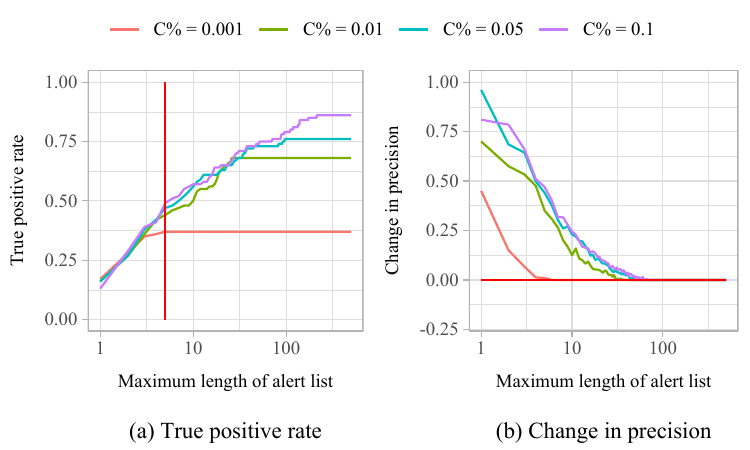}
    \caption{Outlier detection performance under different functional depth thresholds}
    \label{fig:threshold_results}
\end{figure}

\subsubsection{Distribution of outliers across multiple legs} \label{app:sim_outliers}

\begin{figure}[!ht]
    \centering
    \includegraphics[width=\textwidth]{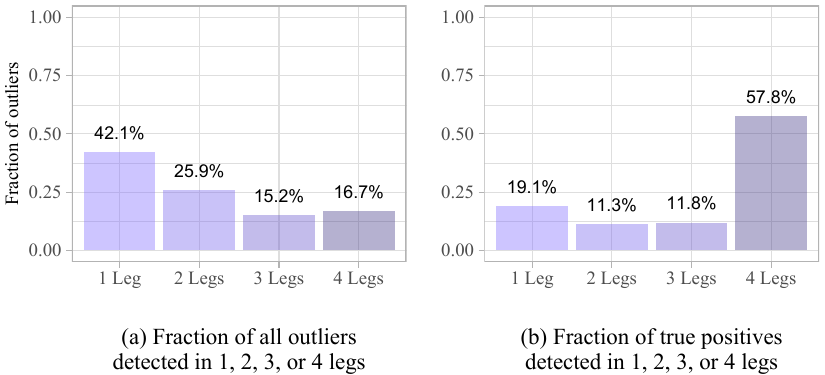}
    \caption{Fraction of outliers detected in 1, 2, 3, or 4 legs}
    \label{fig:sim_num_legs}
\end{figure}

In the scenario where all itineraries are equally affected, a high proportion of outliers should be detected in more than one leg. Figure \ref{fig:sim_num_legs}a illustrates the proportion of outliers detected in 1, 2, 3 or 4 legs: More than half were detected in multiple legs. Figure \ref{fig:sim_num_legs}b shows the proportion of true positives (genuine outliers which were detected), by the number of legs in which they were detected. In contrast with Figure \ref{fig:sim_num_legs}a, a much higher percentage of genuine outliers are detected in all four legs. 

\begin{figure}[!ht]
    \centering
    \includegraphics[width=\textwidth]{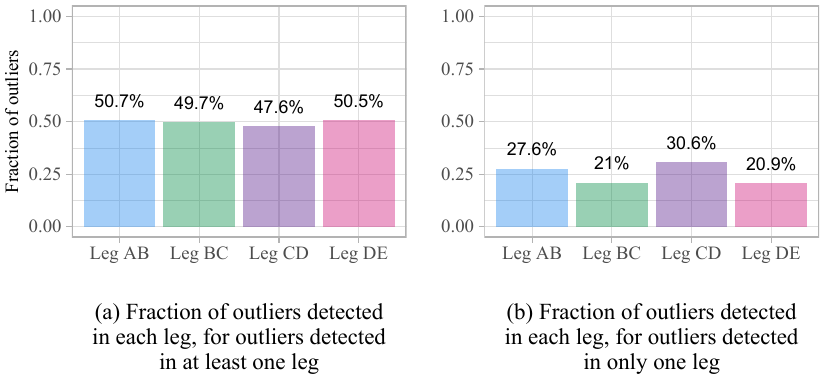}
    \caption{Fraction of outliers detected in each leg}
    \label{fig:sim_leg_prop}
\end{figure}

Given the clustering is correct, we expect an approximately equal number of single-leg outliers in each leg, as shown in Figure \ref{fig:sim_leg_prop}b. If one leg, say DE, had not belonged in this cluster, we would expect a higher proportion of single-leg outliers to have been detected in leg DE. This could be utilised as a method for checking the clustering, after the outlier detection. 

These results motivate aggregating threshold exceedances across legs in two ways: (i) since less than 100\% of genuine outliers were detected in all legs, if outlier detection was carried out only on the leg level, outliers could be missed on some legs. (ii) Given that a much higher proportion of outliers detected in four legs were genuine outliers, by ranking booking patterns detected in all legs as more likely to be outliers, we focus analysts' attention on those more likely to be genuine outliers. 

\subsubsection{False Discovery Rate} 
\label{app:fdr}

\begin{figure}[!ht]
    \centering
    \includegraphics[width=0.6\textwidth]{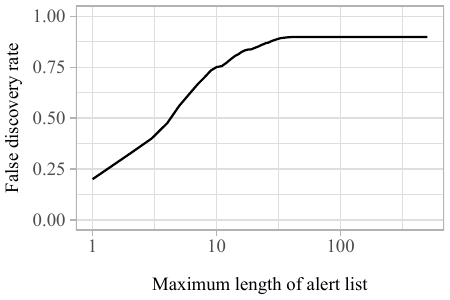}
    \caption{False discovery rate for nonhomogeneous demand-volume outliers}  
	\label{fig:demand_vol_nh_fdr}
\end{figure}

The \textit{false discovery rate (FDR)} is defined as the proportion of booking patterns classified as outliers which were false positives:
\begin{equation}
    FDR = \frac{FP}{TP+FP}
\end{equation}
See Section \ref{sec:outlier_results} for definitions of true and false positives. Figure \ref{fig:demand_vol_nh_fdr} shows the FDR for the case where outlier demand affects all itineraries, and the magnitude is randomly chosen from each of the distributions described in Section \ref{sec:outlier_gen}.

\begin{figure}[!ht]
    \centering
    \includegraphics[width=\textwidth]{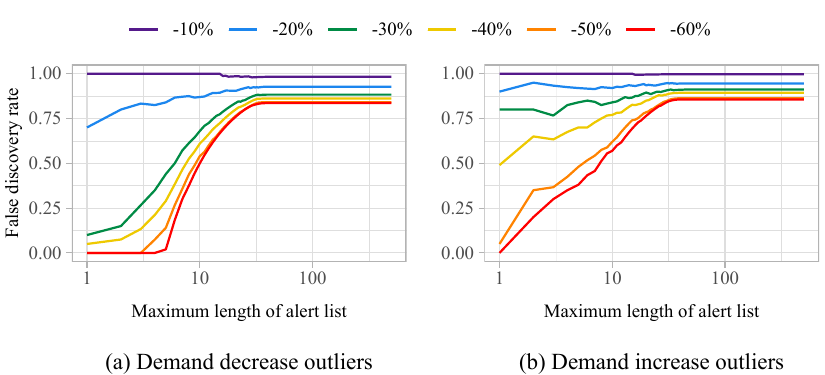}
    \caption{False discovery rate for homogeneous demand-volume outliers by magnitude}
    \label{fig:demand_vol_h_fdr}
\end{figure}

Figure \ref{fig:demand_vol_h_fdr} shows the FDR for each of the magnitudes of outliers considered in the simulation study. Given that smaller magnitude outliers are more similar to the regular demand, these result in higher false discovery rates. 

\subsubsection{Outliers affecting a single itinerary}
\begin{figure}[!ht]
    \centering
    \includegraphics[width=\textwidth]{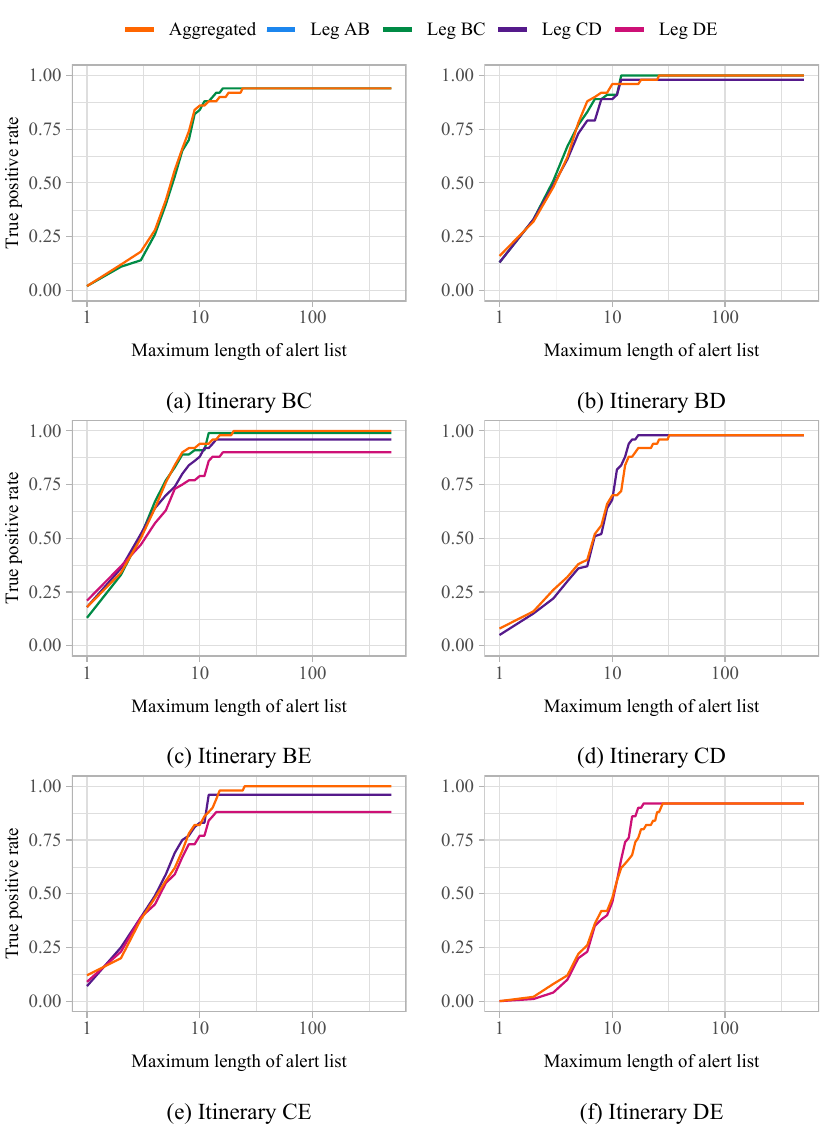}
    \caption{True positive rate for single itinerary outliers (cont.)}
    \label{fig:app_tpr_itin}
\end{figure}

Figure \ref{fig:app_tpr_itin} shows the true positive rate for the remaining itineraries in Figure \ref{fig:tpr_itin} of Section \ref{sec:outlier_results}.
\label{app:single_itin_outliers}

\subsubsection{Outliers affecting a subset of itineraries}
\label{app:subset_itin_outliers}
We consider a case where demand outliers affect only a subset of itineraries. Practical examples for this phenomenon could include trade fairs or conventions as well as regional crises.  In such situations, demand towards (or from) a specific destination is most affected. Here, clustering offers additional benefits in guiding analysts towards those itineraries where they should adjust the forecast or controls.  

We differentiate four scenarios based on the four-leg-network described in Section \ref{sec:simulation}, where events affect demand for itineraries travelling to stations B, C, D, and E respectively. We expect analogous results when customers aim to travel home from events that happened at stations A, B, C, or D respectively, given the symmetry of the demand parameters chosen for the computational study. 

For each of the four possible events considered, we investigate the case where this generates 50\% increase in average leg demand. For simplicity, we assume these passengers are equally split between the itineraries which alight at the relevant station. Table \ref{tab:leg_od_demand_abs} shows the resulting demand increases for each leg.

\begin{table}[!ht]
\resizebox{\textwidth}{!}{\begin{tabular}{c|l|cccc} 
\hline \hline
\multirow{3}{*}{\textbf{\begin{tabular}[c]{@{}c@{}}Event at \\ Station\end{tabular}}} & \multicolumn{1}{c}{\multirow{3}{*}{\textbf{\begin{tabular}[c]{@{}c@{}}Itineraries \\ Affected\end{tabular}}}} & \multicolumn{4}{c}{\textbf{Additional 120 Passengers in Itineraries}} \\
       & \multicolumn{1}{c}{}     & \multicolumn{4}{c}{Resulting Demand Increase per Leg}                     \\ \cline{3-6} & \multicolumn{1}{c}{}          & \textbf{Leg AB}  & \textbf{Leg BC}  & \textbf{Leg CD}  & \textbf{Leg DE}  \\ \hline
B    & A-B   & +120 (+50\%)           & -                & -                & -  \\
C    & A-C, B-C  & +60 (+25\%)   & +120 (+50\%)            & -                & -                \\
D   & A-D, B-D, C-D  & +40 (+16.6\%)          & +80 (+33.3\%)           & +120 (+50\%)          & -                \\
E & A-E, B-E, C-E, D-E   & +30 (12.5\%)          & +60 (+25\%)    & +90 (+37.5\%)         & +120 (+50\%)           \\ \hline \hline
\end{tabular}}
\caption{Changes in leg demand resulting from an additional 120 passengers in itinerary demand}
\label{tab:leg_od_demand_abs}
\end{table}

\begin{figure}[!ht]
    \centering
    \includegraphics[width=\textwidth]{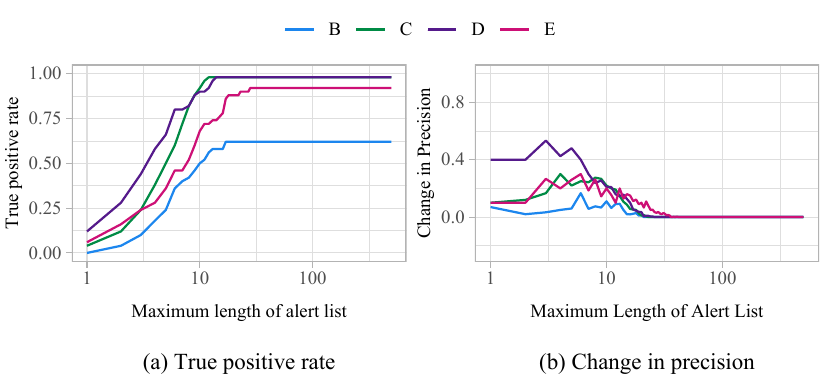}
    \caption{Performance for demand-volume outliers in a subset of itineraries caused by an absolute increase in demand}
    \label{fig:demand_vol_event_abs}
\end{figure}

Figure \ref{fig:demand_vol_event_abs}a shows the true positive rate for each of the cases. Although the event at E generates outliers in more legs, it is not the case that it has the highest true positive rate. This shows that though the approach aggregates across legs, it does not ignore outliers only in a subset of those legs, provided they are sufficiently large. These effects may also be caused by interactions between the booking limits on different legs. For example, in the case of an event at C, large increases in demand in legs AB and BC may cause booking limits to be reached earlier for these legs, which also limits bookings in itineraries such as AD and AE. Hence, an increase in demand for some legs may cause a decrease in bookings for different legs. By jointly considering multiple legs for outlier detection, we are able to detect the knock-on effects of outliers even when the change in demand only affects a subset of legs. The change in precision can be interpreted similarly, in Figure \ref{fig:demand_vol_event_abs}b.

Had we considered outlier detection on a leg-by-leg basis, the outliers were more likely to be missed in some of the legs. By combining information across legs, we are better able to determine which itineraries are affecting the volume of demand. 

\subsubsection{Limit alert list length via outlier severity thresholds} \label{app:outliers_perc}

The results in this paper focus on limiting the length of the ranked alert list simply by the number of alerts it contains as this is most relevant to analysts. However, an alternative approach limits the length of the list by the outlier severity assigned to each departure. For example, classifying a train as an outlier only if its outlier severity is above 80\%.  

\begin{figure}[!ht]
    \centering
    \includegraphics[width=\textwidth]{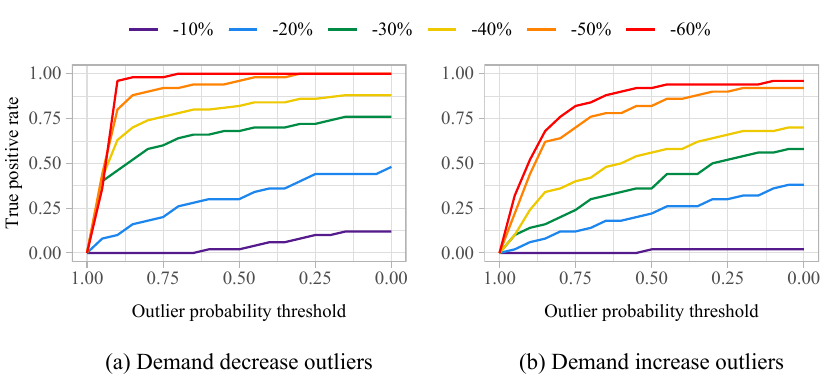}
    \caption{True positive rate for homogeneous demand-volume outliers by magnitude}
    \label{fig:demand_vol_h_tpr_perc}
\end{figure}

\paragraph{Detection results when outliers affect all itineraries} \mbox{} \\
\begin{figure}[!ht]
    \centering
    \includegraphics[width=0.6\textwidth]{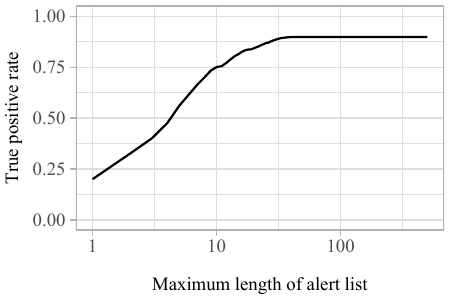}
    \caption{True positive rate for nonhomogeneous demand-volume outliers as minimum outlier severity varies}  
	\label{fig:demand_vol_nh_tpr_perc}
\end{figure}
Figure \ref{fig:demand_vol_nh_tpr_perc} shows the true positive rate as the outlier severity decreases from 100\% to 0\%. Results are similar to those shown in Figure \ref{fig:demand_vol_nh}a. Figure \ref{fig:demand_vol_h_tpr_perc} shows the true positive rate as the outlier severity decreases from 100\% to 0\%, for each magnitude of outlier considered. Results are similar to those shown in Figure \ref{fig:demand_vol_h_tpr}.

\newpage
\paragraph{Detection results when outliers affect a single itinerary} \mbox{} \\
\begin{figure}[!ht]
    \centering
    \includegraphics[width=\textwidth]{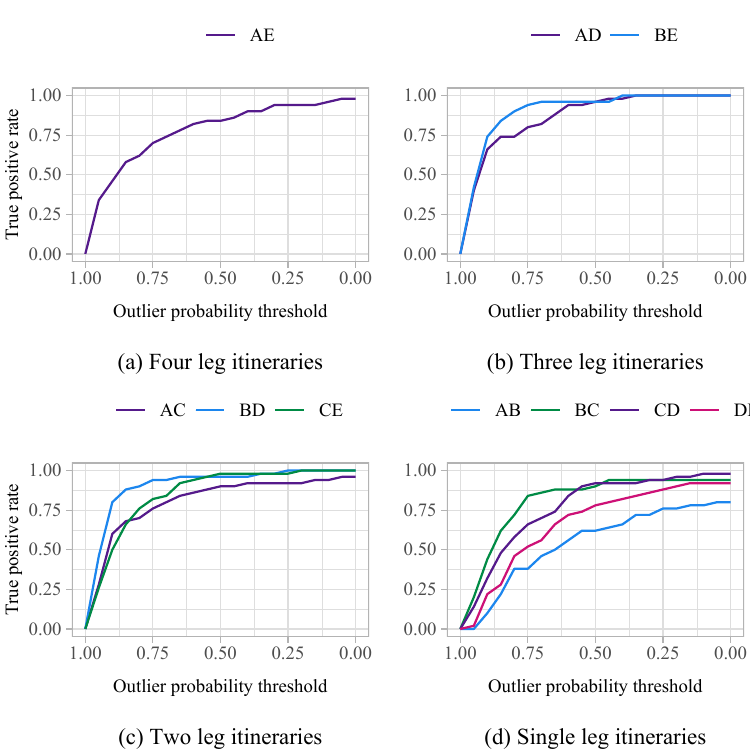}
    \caption{True positive rate for single itinerary demand-volume outliers as minimum outlier severity varies}
    \label{fig:single_itin_tpr_perc}
\end{figure}%

\newpage
\subsection{Detection results when demand is seasonal} \label{app:seasonal}

In this section, we consider the case where demand is non-stationary across departures. In order to make the results of this simulation study more comparable with the other simulation studies included in the paper, we simulate 10 seasonal groups (somewhat analogous to months). We then take 50 "days" from each of the 10 groups to produce 500 booking patterns per simulation, as in the other simulation experiments. All other parameters remain fixed in the simulation.

\begin{figure}[!h]
    \centering
    \includegraphics{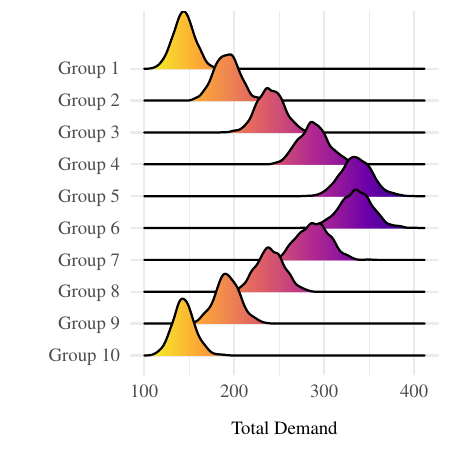}
    \caption{Seasonal demand model showing distributions of total demand for 10 seasonal groups}
    \label{fig:seasonal_demand}
\end{figure}%

We then fit a functional regression model (similar to the model described in Section \ref{sec:db_results} for the empirical study) to remove the seasonality before performing the outlier detection. The quality of the outlier detection will highly depend on the model used to model the seasonal (and trend) components of the demand. A full study of the most suitable approach for modelling this type of demand is outwith the scope of this paper.

\begin{figure}[!ht]
    \centering
    \includegraphics{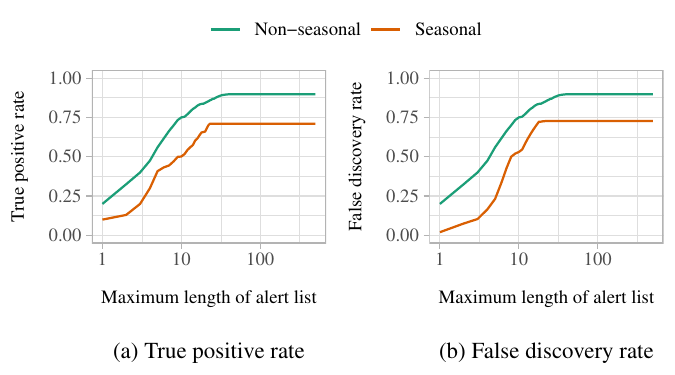}
    \caption{True positive rate and false discovery rate under a seasonal demand model compared to non-seasonal demand}
    \label{fig:seasonal_results}
\end{figure}

In comparison to the non-seasonal demand model, the true positive rate is slightly lower. The false discovery rate is also slightly lower -- see Figure \ref{fig:seasonal_results}. Due to the seasonal variation the overall variance of the demand is higher which causes a lower functional depth threshold (for the same choice of percentile), meaning fewer observations are fall below the threshold.

\newpage
\subsection{Detection results using itinerary-level data} \label{app:itineraries}

The methodology described in this paper focuses on the application to leg-level booking data - since in capacity-based RM only leg-level capacities are required to be stored. Often, the itinerary-level data is never recorded. However, if itinerary level data is available for analysis, one could apply the same approach (either with or without the aggregation step). The results of doing so are presented in this section.

\begin{figure}[!ht]
    \centering
    \includegraphics{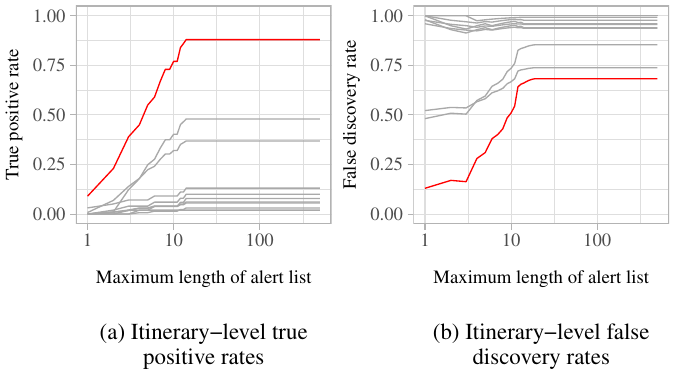}
    \caption{True positive rate and false discovery rate under itinerary level outlier detection}
    \label{fig:itinerary_results}
\end{figure}

Figure \ref{fig:itinerary_results} shows the true positive and false discovery rates from the alert lists generated for each itinerary. It shows that the outlier detection performs well in the most popular itinerary (the itinerary that defines the cluster - A$\rightarrow$E). The results are comparable to running the outlier detection procedure on the leg-level data. In the remaining nine itineraries with far fewer bookings, the outlier detection performs poorly, as there are too few bookings to detect a pattern.

\begin{figure}[!ht]
    \centering
    \includegraphics{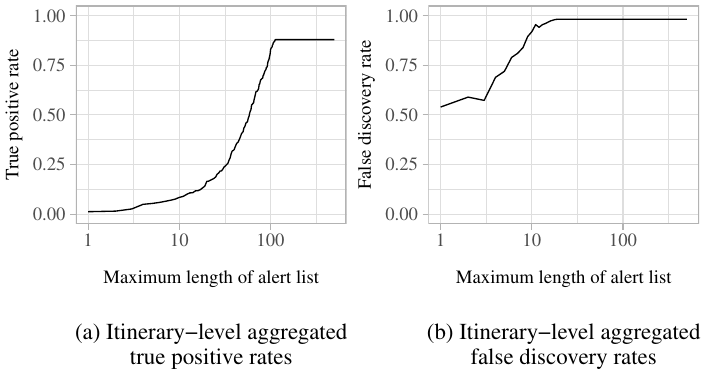}
    \caption{True positive rate and false discovery rate under itinerary level outlier detection with aggregation}
    \label{fig:itinerary_results_agg}
\end{figure}

Figure \ref{fig:itinerary_results_agg} shows that when the itinerary-level outlier detection results are aggregated across itineraries, the overall performance is poorer than the leg-level aggregation. These results suggest that for itinerary-level outlier detection, the better approach would be to identify important itineraries and run the outlier detection routines on those without aggregation. This, however, means that outliers that occur only in a small part of the network (which can cause knock-on effects) would be systematically overlooked. To consider the full network, the leg-based approach that accounts for itineraries through clustering performs better.

\begin{figure}[!ht]
    \centering
    \includegraphics{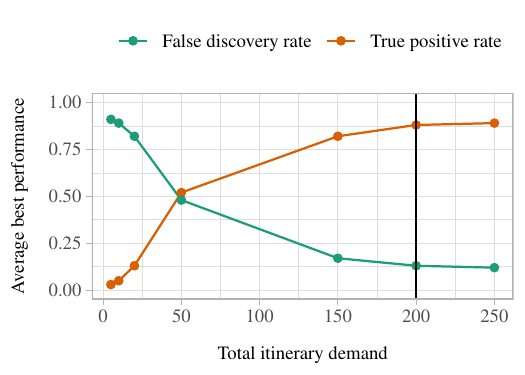}
    \caption{True positive rate and false discovery rate under varying levels of regular demand showing limited success when demand is zero-inflated}
    \label{fig:zeros_results}
\end{figure}

To further illustrate that the issue with itinerary-level analysis is primarily caused by an insufficient number of bookings for less-popular itineraries (making it difficult to detect patterns in the demand), we analyse the performance as the regular itinerary level demand varies. Figure \ref{fig:zeros_results} shows that performance (especially false discovery rate) is poorer when demand is especially low.

\newpage
\subsection{Revenue benefits from forecast adjustments} \label{app:revenue_results}
\begin{figure}[!ht]
    \centering
    \includegraphics[width=\textwidth]{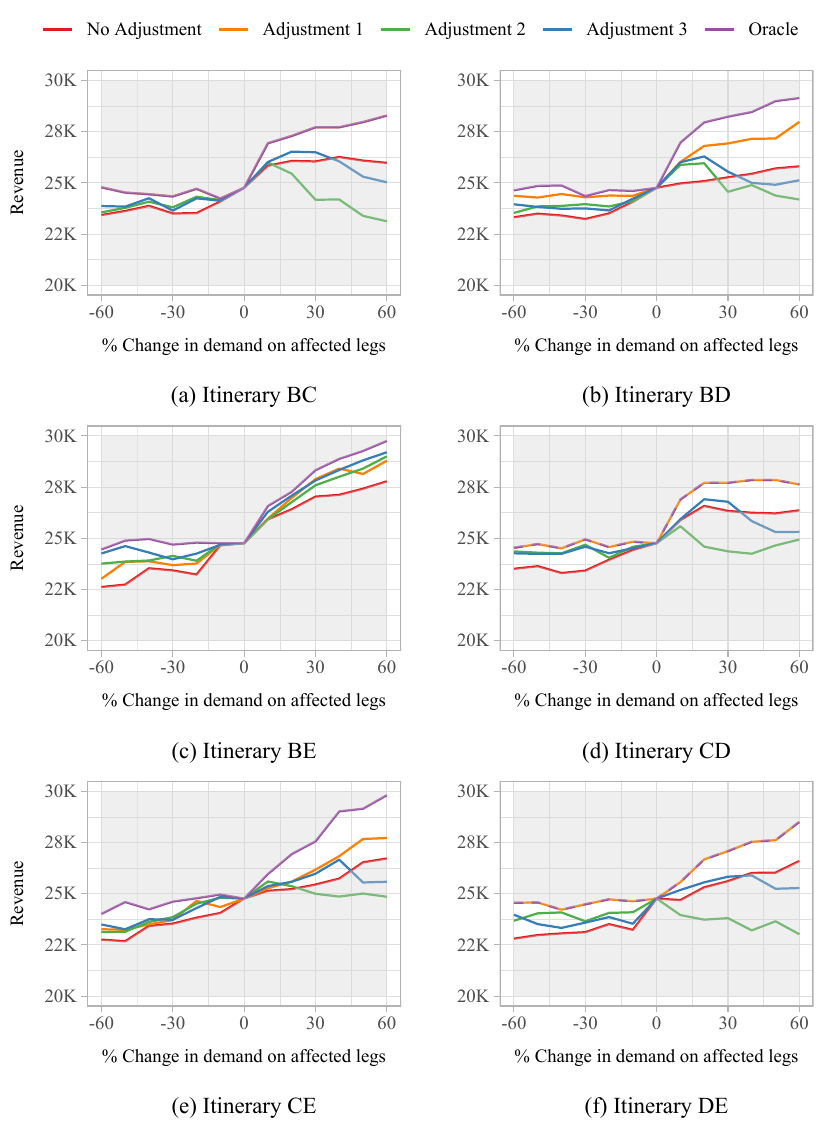}
    \caption{Revenue generated under different itinerary-level forecast adjustments (cont.)}
    \label{fig:app_revenue}
\end{figure}

Figure \ref{fig:app_revenue} shows the true positive rate for the remaining itineraries in Figure \ref{fig:revenue} of Section \ref{sec:revenue_results}.

The analysis in Section \ref{sec:sim_adjustment} constitutes a best-case scenario in which we assume that, if outlier demand affects a particular leg, the outlier is detected in that leg. However, as we show in Section \ref{sec:outlier_results}, even when demand outliers affect multiple legs, the outlier is not always detected in every leg due to noise. Therefore, we additionally compare different adjustments based on the output of the outlier detection, for an outlier in itinerary AE.
\begin{itemize}
    \item \textbf{Adjustment A}: Adjust only the forecasts of the affected single-leg itineraries for those legs in which the outlier is detected. 
    \item \textbf{Adjustment B}: Adjust the forecasts of the affected single-leg itineraries for those legs in which the outlier is detected, \textbf{and} the cluster spanning itinerary (AE). 
\end{itemize}
We compare these both to making no adjustment and to the oracle adjustment. This is still a best-case scenario to some extent, given that we assume the correct magnitude of adjustment is made. 

\begin{figure}[!ht]
    \centering
    \includegraphics[width=\textwidth]{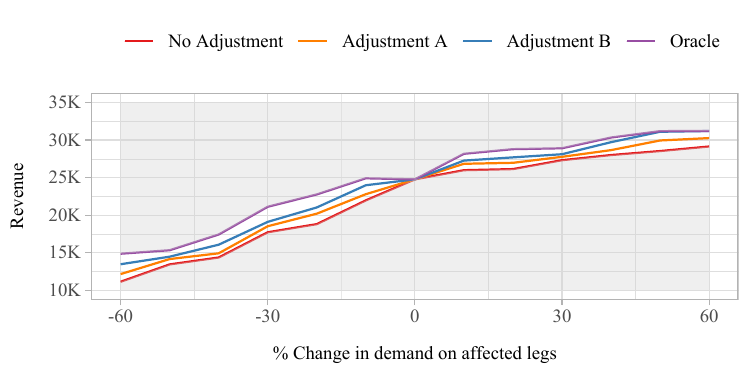}
    \caption{Revenue generated under different forecast adjustments resulting from the outlier detection for outlier demand in itinerary AE}
	\label{fig:revenue_outlier_det}
\end{figure}

Figure \ref{fig:revenue_outlier_det} shows the revenue under adjustments A and B (as described in Section \ref{sec:sim_adjustment}) depending on the output of the outlier detection procedure. Combining adjustments on the leg level with those on the cluster level provides superior results in contrast to leg level adjustments alone. Though making adjustments to only the single-leg itineraries may be risk-averse in the rare cases where an outlier affects only a small subset of the legs within a cluster, it may be detrimental to revenue when outliers affect multiple legs.

\newpage

\newpage

\newpage
\section{Deutsche Bahn booking data} 
\label{sec:db_app}

Appendix \ref{sec:db_app} contains an additional analysis of the empirical booking data from Deutsche Bahn.

\subsection{Model selection for functional regression} 
\label{sec:db_reg}

Due to the functional nature of the data, in order to determine which of the factors result in a better fitting model, we use the \textbf{Cross-Validated Sum Of Integrated Squared Errors} (CV-SSE). 
\begin{equation}
    CV\mbox{-}SSE = \sum_{n=1}^{N} \int (y_{nl}(t) - \hat{y_{nl}(t)} ) dt, 
\end{equation}
where $\hat{y_{nl}(t)}$ is the prediction for the $n^{th}$ booking pattern on the leg $l$, under the model fitted to all but the $n^{th}$ booking pattern. The model which produces the lowest CV-SSE is chosen as the best fitting. Note that unlike other model selection criteria (e.g. AIC), CV-SSE does not take into account the number of parameters. Given that we are not interested in out-of-sample prediction, only in obtaining the best fitting model for our data, over-fitting is not of great concern. The values of the CV-SSE for each of the 12 models considered are shown in Table \ref{tab:model_comp_bookings}.

\begin{table}[htbp]
\centering
\resizebox{\textwidth}{!}{%
\begin{tabular}{r|ccccc|cccc}
\hline \hline 
\multirow{2}{*}{\textbf{Model}}  & \multirow{2}{*}{\textbf{Intercept}} & \multirow{2}{*}{\textbf{Day}} & \multirow{2}{*}{\textbf{Month}} &   \multirow{2}{2.2cm}{\textbf{\centering Short Horizon (I) }}  & \multirow{2}{2.5cm}{\textbf{\centering Short Horizon (C)}}  & \multicolumn{4}{c}{\textbf{CV-SSE}}  \\ 
 &  &  &   &  &  & Leg AB & Leg BC & Leg CD & Leg DE  \\
\hline
\textbf{Model 1} & \checkmark &  &   &  &  & 79974160 & 75034839 & 79529280 & 73824611 \\
\textbf{Model 2} & \checkmark &  &   & \checkmark &  & 58617546  & 52622148 & 52424683  & 50009080 \\
\textbf{Model 3} & \checkmark &  &   &  & \checkmark & 58620898  & 52863263 & 52506946  & 50014984 \\ \hline
\textbf{Model 4} & \checkmark & \checkmark &  &   &  & 27227350  & 35376732 &  32789181 & 30037659\\
\textbf{Model 5} & \checkmark & \checkmark &   & \checkmark &  & 26551341 & 33724380  & 32282900  & 29989390 \\
\textbf{Model 6} & \checkmark & \checkmark &   &  & \checkmark & 26704943 & 34154782  & 32439972  & 30019196 \\ \hline
\textbf{Model 7} & \checkmark &  & \checkmark &  &   & 58620649 & 57895619 & 52638923  & 50015645 \\
\textbf{Model 8} & \checkmark &  & \checkmark & \checkmark &  & 58608640 & 57865403 & 52615801 & 49996331 \\
\textbf{Model 9} & \checkmark &  & \checkmark &   & \checkmark & 58878374 & 57885484 & 52654330 & 50033157 \\ \hline
\textbf{Model 10} & \checkmark & \checkmark & \checkmark &   &  & 24574978 & 25700166 & 21691111 & 21880038 \\
\textbf{Model 11} & \checkmark & \checkmark & \checkmark &  \checkmark &  & \textbf{24519539} & \textbf{25691637} & \textbf{21689686} & \textbf{21878259} \\
\textbf{Model 12} & \checkmark & \checkmark & \checkmark &   & \checkmark & 24546715 & 25697938 & 21724073 & 21896889 \\ \hline \hline 
\end{tabular}}
\caption{Model comparison for functional regression}
\label{tab:model_comp_bookings}
\end{table}

Across all legs, we find that day, month, and shortened booking horizons are all factors that must be taken into account. The inclusion of the days of the week as factors significantly reduces the CV-SSE. In comparison, the inclusion of the booking horizon variable has a smaller, though still positive, effect. We compare two different approaches to accounting for the shortened booking horizon: (i) an indicator function (I) equal to 1 if the booking horizon is shorter, and (ii) a continuous variable (C) between 0 and 1 which gives the length of the shortened horizon as a proportion of the regular length horizon. Based on the CV-SSE scores, shortened booking horizons are best represented by the indicator function i.e. it is important to know that it is shorter but not by how much. The smaller effect of the horizon length variable may be related to the inclusion of the month variable, which is unsurprising given the overlap in the definition of these variables. The values of the CV-SSE are similar for models 2 and 7, where we only consider one month or horizon length as a factor. 

\newpage
\subsection{Residual booking patterns} 
\label{app:db_res}

\begin{figure}[!ht]
    \centering
    \includegraphics[width=\textwidth]{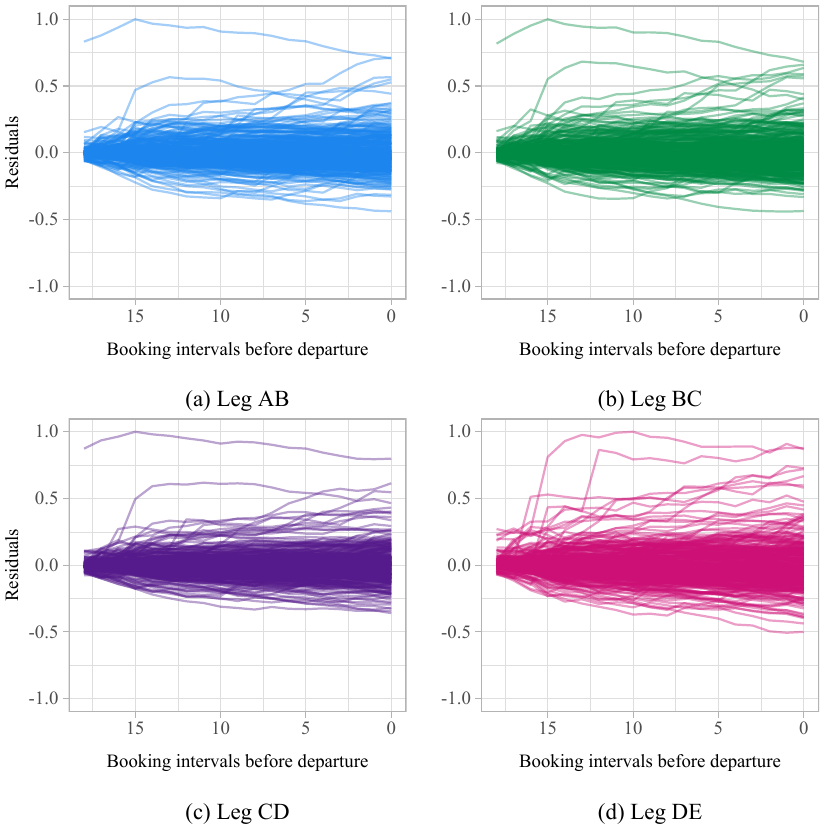}
    \caption{Residual booking patterns}
    \label{fig:db_residuals}
\end{figure}

Figure \ref{fig:db_residuals} shows the residual booking patterns resulting from the functional regression applied in equation~\eqref{eqn:db_funcreg} of Section \ref{sec:db_results}. Compare with Figure \ref{fig:db_bookings} of Section \ref{sec:db_results} -- the obvious outliers are preserved.

\newpage
\subsection{Functional linear regression coefficients}
Table \ref{tab:model_coeffs} shows an example of the coefficients for each explanatory variable over the 19 booking intervals, along with their standard errors. Not all days of the week, or months of the year, are significant throughout the entire booking horizon. Similarly, some days of the week have much stronger significance than others e.g. Saturdays have a much larger impact than Tuesdays. 

\begin{table}[htbp]
\centering
\resizebox{\textwidth}{!}{%
\begin{tabular}{llllllllll}
\hline \hline
\textbf{Coefficient}                                                   & \textbf{t =1}                 & \textbf{2}                    & \textbf{3}                    & \textbf{4}                    & \textbf{5}                    & \textbf{6}                    & \textbf{7}                    & \textbf{8}                    & \textbf{9}                    \\ \hline
$\beta_{0l}$                                                             & 1.4 (1.0)                     & 1.8 (1.1)                     & 5.4 (1.2)                     & 11.2 (1.4)                    & 13.8 (1.5)                    & 14.8 (1.5)                    & 15.2 (1.6)                    & 16.4 (1.6)                    & 17.7 (1.6)                    \\
$\beta_{1l}$ [Mon]                                                       & -0.4 (0.8)                    & -1.0 (0.9)                    & -1.5 (1.0)                    & -2.1 (1.2)                    & -1.6 (1.2)                    & -0.7 (1.3)                    & -0.1 (1.3)                    & 0.5 (1.4)                     & 0.8 (1.4)                     \\
$\beta_{2l}$ [Tue]                                                       & -1.2 (0.8)                    & -2.1 (0.9)                    & -3.7 (1.0)                    & -5.1 (1.2)                    & -5.8 (1.2)                    & -6.3 (1.3)                    & -6.8 (1.2)                    & -6.4 (1.3)                    & -6.4 (1.4)                    \\
$\beta_{3l}$ [Wed]                                                       & -1.2 (0.8)                    & -1.9 (0.9)                    & -3.9 (1.0)                    & -5.2 (1.2)                    & -6.5 (1.2)                    & -7.1 (1.3)                    & -7.3 (1.3)                    & -7.4 (1.3)                    & -8.9 (1.4)                    \\
$\beta_{4l}$ [Thu]                                                       & -1.0 (0.8)                    & -1.3 (0.9)                    & -3.2 (1.0)                    & -4.3 (1.2)                    & -5.6 (1.2)                    & -6.7 (1.3)                    & -6.7 (1.3)                    & -7.1 (1.3)                    & -7.8 (1.4)                    \\
$\beta_{5l}$ [Fri]                                                       & -0.5 (0.8)                    & -0.4 (0.9)                    & -0.7 (1.0)                    & -0.3 (1.2)                    & 0.8 (1.2)                     & 1.5 (1.3)                     & 2.6 (1.3)                     & 2.7 (1.4)                     & 2.9 (1.4)                     \\
$\beta_{6l}$ [Sat]                                                       & 5.5 (0.8)                     & 10.9 (0.9)                    & 17.1 (1.1)                    & 25.5 (1.2)                    & 30.4 (1.3)                    & 33.5 (1.3)                    & 36.9 (1.3)                    & 38.6 (1.4)                    & 41.2 (1.4)                    \\
$\beta_{7l}$ [Jan]                                                       & -0.7 (0.9)                    & -0.3 (1.0)                    & -1.3 (1.2)                    & -4.3 (1.3)                    & -5.0 (1.4)                    & -4.8 (1.5)                    & -4.5 (1.5)                    & -5.0 (1.5)                    & -4.4 (1.6)                    \\
$\beta_{8l}$ [Feb]                                                       & 0.3 (1.0)                     & 1.8 (1.1)                     & 0.0 (1.3)                     & -4.0 (1.4)                    & -3.5 (1.5)                    & -2.3 (1.6)                    & -0.3 (1.6)                    & 0.3 (1.7)                     & 1.6 (1.7)                     \\
$\beta_{9l}$ [Mar]                                                       & 0.2 (1.0)                     & 0.9 (1.1)                     & 0.4 (1.2)                     & -3.2 (1.4)                    & -4.4 (1.5)                    & -4.0 (1.6)                    & -3.2 (1.6)                    & -2.9 (1.7)                    & -2.5 (1.7)                    \\
$\beta_{10l}$ [Apr]                                                      & \multicolumn{1}{l}{0.3 (1.0)} & \multicolumn{1}{l}{2.1 (1.1)} & \multicolumn{1}{l}{1.2 (1.3)} & \multicolumn{1}{l}{0.9 (1.4)} & \multicolumn{1}{l}{3.1 (1.5)} & \multicolumn{1}{l}{4.8 (1.6)} & \multicolumn{1}{l}{5.6 (1.6)} & \multicolumn{1}{l}{5.6 (1.7)} & \multicolumn{1}{l}{5.9 (1.7)} \\
$\beta_{11l}$ [May]                                                      & 3.5 (1.0)                     & 4.4 (1.1)                     & 4.1 (1.2)                     & 0.7 (1.4)                     & -0.1 (1.5)                    & 0.0 (1.6)                     & 0.6 (1.6)                     & 0.5 (1.7)                     & 0.9 (1.7)                     \\
$\beta_{12l}$ [Jun]                                                      & 0.2 (1.0)                     & 1.9 (1.1)                     & 0.7 (1.3)                     & -1.9 (1.4)                    & -2.6 (1.5)                    & -2.2 (1.6)                    & -1.9 (1.6)                    & -2.0 (1.7)                    & -1.5 (1.7)                    \\
$\beta_{13l}$ [Jul]                                                      & 0.4 (0.8)                     & 1.6 (1.1)                     & 1.4 (1.2)                     & -1.5 (1.4)                    & -1.8 (1.5)                    & -0.7 (1.6)                    & -0.3 (1.6)                    & 0.1 (1.7)                     & 0.9 (1.7)                     \\
$\beta_{14l}$ [Aug]                                                      & 1.5 (1.0)                     & 3.3 (1.1)                     & 3.3 (1.3)                     & 3.7 (1.5)                     & 5.0 (1.6)                     & 6.4 (1.6)                     & 7.7 (1.7)                     & 8.3 (1.7)                     & 9.8 (1.7)                     \\
$\beta_{15l}$ [Sep]                                                      & 1.9 (1.1)                     & 2.5 (1.2)                     & 2.1 (1.3)                     & -1.2 (1.5)                    & -0.7 (1.6)                    & 0.6 (1.6)                     & 1.5 (1.7)                     & 2.3 (1.7)                     & 3.5 (1.8)                     \\
$\beta_{16l}$ [Oct]                                                      & 1.6 (1.0)                     & 3.5 (1.1)                     & 3.8 (1.2)                     & 4.7 (1.4)                     & 7.0 (1.5)                     & 8.8 (1.6)                     & 11.2 (1.6)                    & 12.6 (1.7)                    & 15.5 (1.7)                    \\
$\beta_{17l}$ [Nov]                                                      & 1.0 (0.9)                     & 2.8 (1.1)                     & 4.2 (1.3)                     & 4.5 (1.4)                     & 6.7 (1.5)                     & 9.3 (1.6)                     & 11.0 (1.6)                    & 11.5 (1.7)                    & 12.5 (1.7)                    \\
$\beta_{18l}$ [Horizon] & -1.2 (0.9)                    & -0.2 (1.0)                    & 0.7 (1.2)                     & -0.8 (1.3)                    & -1.1 (1.4)                    & -0.6 (1.5)                    & -0.5 (1.5)                    & -0.2 (1.5)                    & -0.4 (1.6)                    \\  \hline
\multicolumn{1}{l}{}                                                   & \multicolumn{1}{l}{}          & \multicolumn{1}{l}{}          & \multicolumn{1}{l}{}          & \multicolumn{1}{l}{}          & \multicolumn{1}{l}{}          & \multicolumn{1}{l}{}          & \multicolumn{1}{l}{}          & \multicolumn{1}{l}{}          & \multicolumn{1}{l}{}          \\
\multicolumn{1}{l}{}                                                   & \multicolumn{1}{l}{}          & \multicolumn{1}{l}{}          & \multicolumn{1}{l}{}          & \multicolumn{1}{l}{}          & \multicolumn{1}{l}{}          & \multicolumn{1}{l}{}          & \multicolumn{1}{l}{}          & \multicolumn{1}{l}{}          & \multicolumn{1}{l}{}          \\ \hline
\textbf{t = 10}                                                        & \textbf{11}                   & \textbf{12}                   & \textbf{13}                   & \textbf{14}                   & \textbf{15}                   & \textbf{16}                   & \textbf{17}                   & \textbf{18}                   & \textbf{19}                   \\ \hline
18.5 (1.7)                                                             & 18.7 (1.7)                    & 19.5 (1.7)                    & 20.7 (1.8)                    & 22.2 (1.8)                    & 24.1 (1.8)                    & 26.0 (1.8)                    & 27.9 (1.9)                    & 29.0 (1.9)                    & 30.4 (2.0)                    \\
1.1 (1.4)                                                              & 1.1 (1.4)                     & 1.1 (1.5)                     & 0.4 (1.5)                     & -0.6 (1.5)                    & -1.1 (1.5)                    & -2.5 (1.6)                    & -3.7 (1.6)                    & -3.5 (1.6)                    & -2.1 (1.7)                    \\
-6.8 (1.4)                                                             & -5.8 (1.4)                    & -5.7 (1.5)                    & -6.4 (1.5)                    & -7.4 (1.5)                    & -8.0 (1.5)                    & -9.3 (1.6)                    & -10.2 (1.6)                   & -10.0 (1.6)                   & -9.1 (1.6)                    \\
-9.2 (1.4)                                                             & -8.9 (1.4)                    & -9.1 (1.5)                    & -10.0 (1.5)                   & -11.2 (1.5)                   & -12.7 (1.5)                   & -14.5 (1.6)                   & -14.7 (1.6)                   & -15.1 (1.6)                   & -15.2 (1.7)                   \\
-7.8 (1.4)                                                             & -7.6 (1.4)                    & -7.7 (1.5)                    & -8.9 (1.5)                    & -10.7 (1.5)                   & -12.1 (1.5)                   & -12.9 (1.6)                   & -13.6 (1.6)                   & -13.7 (1.6)                   & -13.7 (1.7)                   \\
4.2 (1.4)                                                              & 5.0 (1.4)                     & 5.4 (1.5)                     & 4.3 (1.5)                     & 4.0 (1.5)                     & 4.6 (1.5)                     & 5.0 (1.6)                     & 5.8 (1.6)                     & 6.4 (1.6)                     & 7.1 (1.7)                     \\
45.0 (1.4)                                                             & 46.1 (1.5)                    & 48.1 (1.5)                    & 49.4 (1.5)                    & 50.7 (1.5)                    & 53.1 (1.6)                    & 55.1 (1.6)                    & 57.1 (1.6)                    & 58.5 (1.7)                    & 59.4 (1.7)                    \\
-3.7 (1.6)                                                             & -3.0 (1.7)                    & -3.8 (1.7)                    & -3.6 (1.7)                    & -3.1 (1.7)                    & -3.5 (1.8)                    & -3.3 (1.8)                    & -3.8 (1.8)                    & -3.1 (1.9)                    & -3.8 (1.9)                    \\
3.3 (1.8)                                                              & 3.7 (1.8)                     & 3.9 (1.8)                     & 4.9 (1.8)                     & 5.9 (1.9)                     & 6.8 (1.9)                     & 7.8 (1.9)                     & "8.2 (2.0)                    & 9.5 (2.0)                     & 9.2 (2.1)                     \\
-1.6 (1.7)                                                             & -0.8 (1.8)                    & -1.1 (1.8)                    & -0.6 (1.8)                    & -0.2 (1.9)                    & -0.6 (1.9)                    & -0.4 (1.9)                    & -0.8 (2.0)                    & -0.3 (2.0)                    & -0.7 (2.0)                    \\
6.1 (1.8)                                                              & \multicolumn{1}{l}{6.3 (1.8)} & \multicolumn{1}{l}{6.3 (1.8)} & \multicolumn{1}{l}{6.8 (1.8)} & \multicolumn{1}{l}{7.2 (1.9)} & \multicolumn{1}{l}{7.3 (1.9)} & \multicolumn{1}{l}{7.2 (1.9)} & \multicolumn{1}{l}{7.1 (2.0)} & 8.3 (2.0)                     & 8.8 (2.0)                     \\
1.1 (1.7)                                                              & 1.4 (1.8)                     & 1.3 (1.8)                     & 1.8 (1.8)                     & 2.5 (1.9)                     & 2.3 (1.9)                     & 2.4 (1.9)                     & 2.1 (2.0)                     & 3.1 (2.0)                     & 3.5 (2.0)                     \\
-1.0 (1.8)                                                             & -0.4 (1.8)                    & -0.2 (1.8)                    & 0.9 (1.8)                     & 1.5 (1.9)                     & 1.1 (1.9)                     & 1.1 (1.9)                     & 0.7 (2.0)                     & 2.0 (2.0)                     & 2.9 (2.0)                     \\
1.8 (1.7)                                                              & 2.5 (1.8)                     & 2.6 (1.8)                     & 3.1 (1.8)                     & 4.0 (1.9)                     & 4.0 (1.9)                     & 4.4 (1.9)                     & 4.6 (2.0)                     & 6.3 (2.0)                     & 6.9 (2.0)                     \\
11.6 (1.8)                                                             & 12.8 (1.8)                    & 13.2 (1.8)                    & 14.0 (1.9)                    & 15.4 (1.9)                    & 16.4 (1.9)                    & 17.1 (2)                      & 17.4 (2.0)                    & 19.2 (2.1)                    & 21.7 (2.1)                    \\
5.1 (1.8)                                                              & 6.8 (1.8)                     & 7.3 (1.9)                     & 8.7 (1.9)                     & 9.9 (1.9)                     & 10.2 (2.0)                    & 11.6 (2.0)                    & 11.6 (2.0)                    & 12.4 (2.1)                    & 13.3 (2.1)                    \\
17.0 (1.7)                                                             & 18.4 (1.8)                    & 19.2 (1.8)                    & 20.7 (1.8)                    & 22.0 (1.9)                    & 22.9 (1.9)                    & 24.6 (1.9)                    & 25.3 (2.0)                    & 27.6 (2.0)                    & 28.0 (2.0)                    \\
13.4 (1.8)                                                             & 14.0 (1.8)                    & 14.0 (1.8)                    & 15.1 (1.8)                    & 15.8 (1.9)                    & 15.9 (1.9)                    & 16.2 (1.9)                    & 16.1 (2.0)                    & 16.6 (2.0)                    & 16.2 (2.0)                    \\
-0.2 (1.6)                                                             & 0.2 (1.7)                     & 1.2 (1.7)                     & 1.9 (1.7)                     & 2.6 (1.7)                     & 2.9 (1.8)                     & 3.2 (1.8)                     & 4.5 (1.8)                     & 5.4 (1.9)                     & 6.0 (1.9)                    
  \\ \hline \hline
\end{tabular}}
\caption{Functional linear regression coefficients (and standard errors)}
\label{tab:model_coeffs}
\end{table}

\newpage
\subsection{Functional depths} \label{sec:db_depth}
\begin{figure}[!ht]
    \centering
    \includegraphics[width=\textwidth]{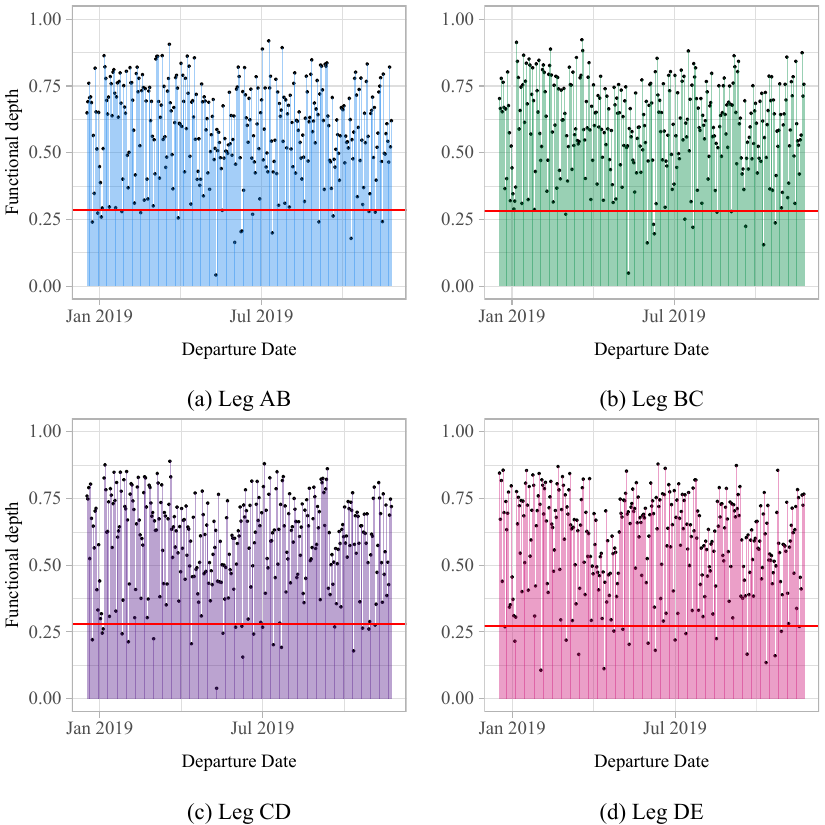}
    \caption{Functional depths}
    \label{fig:db_depth}
\end{figure}

Figure \ref{fig:db_depth} shows the functional depths for the empirical residual booking patterns, before the functional depths are transformed into the $z_{nl}$, as shown in Figure \ref{fig:db_diffs} of Section \ref{sec:db_results}.

\newpage
\subsection{Probability plots for GPD and Exponential distributions}  \label{app:pp_plots}
\begin{figure}[!ht]
    \centering
    \includegraphics[width=\textwidth]{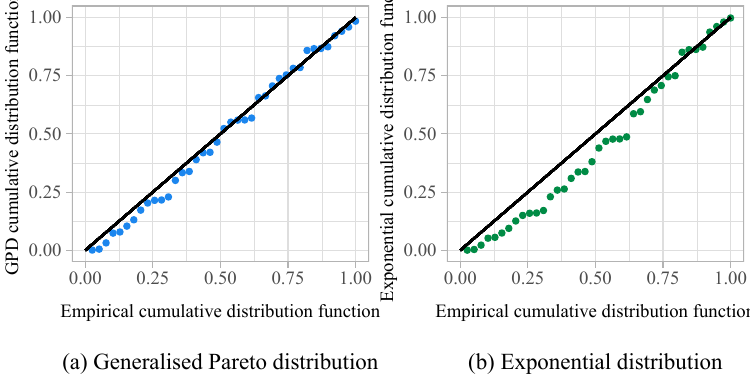}
    \caption{P-P plots}
    \label{fig:pp_plots}
\end{figure}

Given that, if both \(\mu = 0\) and \(\xi = 0\), the GPD reduces to an exponential distribution, it is appropriate to compare the fit of the GPD with an exponential distribution to check if the inclusion of additional parameters is beneficial. Figure \ref{fig:pp_plots} shows the P-P plots, i.e. the fitted theoretical CDF against the empirical CDF for the GPD (Figure \ref{fig:pp_plots}a) and the Exponential distribution (Figure \ref{fig:pp_plots}b). The GPD provides a closer fit to the empirical data and the additional parameters better account for the shape of the distribution. 
The GPD does not provide a perfect fit, with the probabilities in the bottom left of Figure \ref{fig:pp_plots}a on consistently being underestimated. However, given that we assume points with very low probability are more likely to be false positives, under-estimating may actually be beneficial. Further, only the highly-ranked outliers i.e. those with high probability, are likely to be considered by an analyst due to time constraints. The GPD provides a very good fit for those data points. If there is a sufficiently large number of threshold exceedances, an empirical distribution could alternatively be used to compute the probabilities. 

\subsection{Distribution of outliers across multiple legs} \label{app:db_outliers}
\begin{figure}[!ht]
    \centering
    \includegraphics[width=0.5\textwidth]{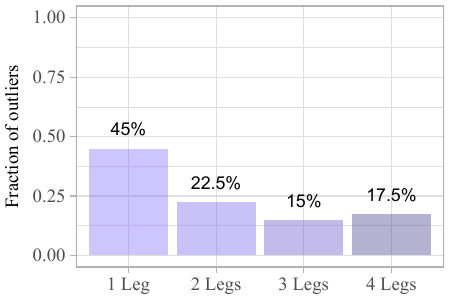} 
    \caption{Fraction of all outliers detected in 1, 2, 3, or 4 legs}
	\label{fig:db_num_legs}
\end{figure}

The proportion of outliers found in each number of legs is shown in Figure \ref{fig:db_num_legs}, with over half of the outliers detected in multiple legs. Compared with Figure \ref{fig:sim_num_legs}, this shows a similar proportion of outliers as found in the simulation study. 

\begin{figure}[!ht]
    \centering
    \includegraphics[width=\textwidth]{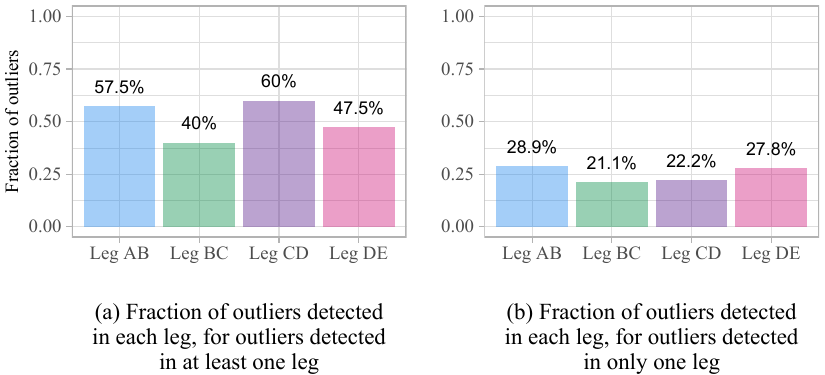} 
    \caption{Fraction of outliers detected in each leg}
    \label{fig:db_leg_prop}
\end{figure}

Figure \ref{fig:db_leg_prop}a shows the proportion of total outlying booking patterns in terms of which legs they were detected as outliers in. Figure \ref{fig:db_leg_prop}b shows the proportion in each leg of outlying booking patterns detected in one leg only. The proportions are fairly evenly split between the different legs. This reassures us that the correct clustering was chosen - if leg DE did in fact belong to a separate second cluster, we would expect a higher proportion of single-leg outliers to have been found in leg DE -- compare with Figure \ref{fig:sim_leg_prop}.


\subsection{Simulation verification} \label{app:sim_ver}
\begin{figure}[!h]
    \centering
    \includegraphics[width=\textwidth]{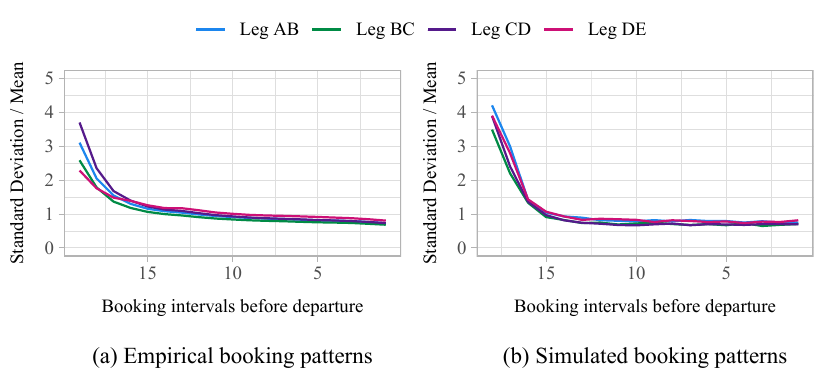}
    \caption{Comparison of standard deviation divided by mean of booking patterns}
    \label{fig:sd_mean}
\end{figure}

In order to validate the parameter choices used to simulate booking patterns, we compare the resulting simulated booking patterns with the empirical booking patterns. We consider the standard deviation and mean of the bookings across the booking horizon of each in Figure \ref{fig:sd_mean}. Both the empirical and simulated booking patterns show a similar shape and magnitude of the relationship between the mean and standard deviation across the booking horizon. 

We also compare the correlations between the different legs for both the empirical and simulated data. Table \ref{tab:db_corr} shows the functional dynamical correlation between the empirical booking patterns, and empirical residual booking patterns, for each leg. Table \ref{tab:sim_leg_corr} shows the corresponding correlations between the simulated booking patterns. The values are similar and the rate of decay between legs as they get further apart follows a similar pattern.

\begin{table}[!htb]
    \begin{subtable}{.47\linewidth}
      \centering
        \resizebox{\textwidth}{!}{\begin{tabular}{c|cccc}
        \hline \hline
        & \textbf{Leg AB} & \textbf{Leg BC} & \textbf{Leg CD} & \textbf{Leg DE} \\ \hline
        \textbf{Leg AB} & - & 0.95 & 0.83 & 0.70 \\ 
        \textbf{Leg BC} & - & - & 0.83 & 0.66 \\ 
        \textbf{Leg CD} & - & - & - & 0.78 \\ 
        \textbf{Leg DE} & - & - & - & - \\ 
        \hline \hline
        \end{tabular}}
    \caption{Booking patterns}
    \label{tab:corr_bookings}
    \end{subtable}%
    \hspace{0.2cm}
    \begin{subtable}{.47\linewidth}
      \centering
        \resizebox{\textwidth}{!}{\begin{tabular}{c|cccc}
        \hline \hline
        & \textbf{Leg AB} & \textbf{Leg BC} & \textbf{Leg CD} & \textbf{Leg DE} \\ \hline
        \textbf{Leg AB} & - & 0.92 & 0.75 & 0.58 \\ 
        \textbf{Leg BC} & - & - & 0.88 & 0.74 \\ 
        \textbf{Leg CD} & - & - & - & 0.84 \\ 
        \textbf{Leg DE} & - & - & - & - \\ 
        \hline \hline
        \end{tabular}}
    \caption{Residual booking patterns}
    \label{tab:corr_residuals}
    \end{subtable} 
    \caption{Functional dynamical correlation of empirical booking patterns}
    \label{tab:db_corr}
\end{table}

\begin{table}[!ht]
    \centering
\begin{tabular}{c|cccc}
\hline \hline
  & \textbf{Leg AB} & \textbf{Leg BC} & \textbf{Leg CD} & \textbf{Leg DE} \\ \hline
\textbf{Leg AB} & - & 0.81 & 0.72 & 0.60 \\ 
\textbf{Leg BC} & - & - & 0.86 & 0.68 \\ 
\textbf{Leg CD} & - & - & - & 0.78 \\ 
\textbf{Leg DE} & - & - & - &  -    \\ \hline \hline
\end{tabular}
\caption{Functional dynamical correlation of simulated booking patterns }
\label{tab:sim_leg_corr}
\end{table}

\end{document}